\documentclass[reprint,amsmath,amssymb,aps,prd]{revtex4-1}
\usepackage{graphics}
\begin{document}
\title{Scale-dependent bias of galaxies and $\mu$-type distortion of
the cosmic microwave background spectrum from single-field inflation
with a modified initial state}
\author{Jonathan Ganc}
\affiliation{Texas Cosmology Center and the Department of Physics,
The University of Texas at Austin, 1 University Station, C1600, Austin,
TX 78712}
\author{Eiichiro Komatsu}
\affiliation{Texas Cosmology Center and the Department of Astronomy,
The University of Texas at Austin, 1 University Station, C1400, Austin,
TX 78712}
\affiliation{Kavli Institute for the Physics and Mathematics of the
Universe, Todai Institutes for Advanced Study, the University of Tokyo,
Kashiwa, Japan 277-8583 (Kavli IPMU, WPI)}
\affiliation{Max Planck Institut f\"ur Astrophysik, Karl-Schwarzschild-Str. 1, 85741 Garching, Germany}
\newcommand{\mbf}[1]{\mathbf{#1}}
\newcommand{\mpl}[0]{M_{\text{pl}}}
\begin{abstract}
We investigate the phenomenological consequences of a modification of
 the initial state of quantum fluctuations of a single inflationary
 field. While single-field inflation with the standard Bunch-Davies
 initial vacuum state does not generally produce a measurable
 three-point function (bispectrum) in the so-called squeezed triangle
 configuration (where one wavenumber, $k$, is much smaller than the
 other two, $k\ll k_1\approx k_2$), allowing for a non-standard initial
 state produces an exception. Here, we calculate the signature of an
 initial state modification in single-field slow-roll inflation as it
 would appear in both the scale-dependent bias of the large-scale
 structure (LSS) and $\mu$-type distortion in the black-body spectrum of
 the cosmic microwave background (CMB). We parametrize the initial state
 modifications and identify certain choices of parameters as natural,
 though we also note some fine-tuned choices that can yield a larger
 bispectrum. In both cases, we observe a distinctive $k^{-3}$ signature
 in LSS (as opposed to the $k^{-2}$ of the so-called local-form
 bispectrum). As a non-zero bispectrum in the squeezed configuration
 correlates one long-wavelength mode with two short-wavelength modes, it
 induces a correlation between the CMB temperature anisotropy observed
 on large scales with the temperature-anisotropy-squared on very small
 scales; this correlation persists as the small-scale anisotropy-squared
 is processed into the $\mu$-type distortion of the black-body
 spectrum. While the correlation induced by the local-form bispectrum
 turns out to be too small to detect in near future, a modified initial
 vacuum state enhances the signal by a large factor owing to an extra
 factor of $k_1/k$ compared to the local form. For example, a proposed
 absolutely-calibrated experiment, PIXIE, is expected to detect this
 correlation with a signal-to-noise ratio greater than 10, for an
 occupation number of about $0.5$ in the observable modes. Relatively-calibrated experiments such as Planck and LiteBIRD should also be able
 to measure this effect, provided that the relative calibration between
 different frequencies meets the required precision. Our study suggests
 that the CMB anisotropy, the distortion of the CMB black-body spectrum,
 and the large-scale structure of the universe offer new ways to probe
 the initial state of quantum fluctuations.
\end{abstract}
\maketitle
\section{Introduction}
\label{sec:intro}
While cosmologists have accumulated extensive evidence for an
early-universe inflationary period, the cause and dynamical specifics of
that epoch remain unclear. Current and upcoming measurements will
provide increasingly precise measurements of the effects of inflation,
demanding that theorists persist in relating these observations to
inflation's underlying mechanism. Primordial non-Gaussianity is a
popular discriminant among the proposed models of inflation (e.g.,
\cite{bartolo/etal:2004,2009astro2010S.158K,Chen:2010xka}). 

The scalar curvature perturbation, $\zeta$, which appears in
the space-space part of the metric in a suitable gauge as
$g_{ij}=a^2(t)e^{2\zeta}\delta_{ij}$ (where $a(t)$ is the
Robertson-Walker scale factor), is a convenient quantity relating the
observables such as the cosmic microwave background (CMB) and the
large-scale structure (LSS) of the universe to the primordial
perturbations generated during inflation. In particular, this quantity
is conserved outside the horizon for single-field inflation (e.g.,
\cite{Weinberg:2003}). We shall define the two-point function (power
spectrum, denoted as $P_\zeta(k)$) and the three-point function
(bispectrum, denoted as $B_\zeta(k_1,k_2,k)$) of $\zeta$ in
Fourier space as follows:
\begin{eqnarray}
 \langle\zeta_{\mbf{k}_1}\zeta_{\mbf{k}}\rangle&=&(2\pi)^3\delta(\mbf{k}_1+\mbf{k})P_\zeta(k),\\
 \langle\zeta_{\mbf{k}_1}\zeta_{\mbf{k}_2}\zeta_{\mbf{k}}\rangle&=&(2\pi)^3\delta(\mbf{k}_1+\mbf{k}_2+\mbf{k})B_\zeta(k_1,k_2,k).
\end{eqnarray}
The current data constrain the shape of $P_\zeta(k)$ as
$P_\zeta(k)\propto k^{n_s-4}$ with $n_s=0.96\pm 0.01$ \cite{Komatsu:2010fb,Dunkley:2010ge,*Keisler:2011aw}.

The so-called local-form bispectrum defined as
\cite{gangui/etal:1994,*verde/etal:2000,Komatsu:2001rj}
\begin{equation}
 B^{\rm  local}_\zeta(k_1,k_2,k)\equiv \frac65f_{\rm NL}\left[
P_\zeta(k_1)P_\zeta(k_2)+\mbox{(2 perm.)}
\right],
\label{eq:localform}
\end{equation}
is particularly interesting, both because a detection of the
primordial bispectrum at the level of $f_{\rm NL}\gg 1$ would disfavor
single-field inflation \cite{Malda,Acquaviva:2002ud,creminelli04,Creminelli:2011rh} and
because it is easy to measure the primordial signal since few late-time
effects can produce the local-form bispectrum.
The most important
contamination of $f_{\rm NL}$ known to date is due to the lensing-Integrated
Sachs-Wolfe (ISW) effect bispectrum
\cite{Goldberg:1999xm,*Verde:2002mu,*Smith:2006ud,*Serra:2008wc,*Hanson:2009kg,*Junk:2012qt},
which can be calculated precisely and removed. The contamination of
$f_{\rm NL}$ due to non-linearity in the photon-baryon fluid has been
shown to be at most one
\cite{Nitta:2009jp,*Creminelli:2011sq,*Bartolo:2011wb}.

The local-form bispectrum has the largest signal in the
so-called ``squeezed triangle configuration,'' for which one of the
wavenumbers, say, $k$, is much smaller than the other two, $k\ll
k_1\approx k_2$. This can be seen from Eq.~(\ref{eq:localform}): as
$P_\zeta(k)\propto k^{-3}$ for a scale-invariant spectrum ($n_s=1$), the
bispectrum is maximized when $k$ is taken to be small. In this
limit, one finds:
\begin{equation}
 B^{\rm local}(k_1,k_1,k\to 0) \to \frac{12}{5}f_{\rm
  NL}P_\zeta(k_1)P_\zeta(k)\propto \frac1{k_1^3k^3},
\label{eq:locallimit}
\end{equation}
for a scale-invariant spectrum.

Recently, Agullo and Parker have shown that a non-standard
initial state of quantum fluctuations generated during single-field
inflation can 
enhance the bispectrum in the squeezed configuration by a factor of
$k_1/k$, i.e., $B(k_1,k_1,k\to 0)\propto 1/(k_1^2k^4)$
\cite{Agullo:2010ws}. This would have profound implications for
observations of the bispectrum in the squeezed configuration. For
example, the signature in the bispectrum of CMB of this
model was investigated in a paper by one of the authors
\cite{Ganc:2011dy}, who found that the
model could produce a measurable local $f_{\text{NL}}$ signal in the CMB.  

The primordial bispectrum in the squeezed configuration
was initially constrained mostly by measurements of the temperature
anisotropy of the CMB
\cite{Komatsu:2001wu,*Komatsu:2003fd,Komatsu:2010fb}. However, over
time, tools for observing the bispectrum have proliferated,
providing a variety of ways to compare inflationary models. In this
paper, we will explore two such methods: 
\begin{itemize}
 \item [1.] In the large scale-structure (LSS) of the
       universe, the local-form bispectrum leaves a
       signature by contributing a scale-dependence to the halo bias, $b(k)$
       \cite{Dalal:2007cu,Slosar:2008hx,Matarrese:2008nc}. For
       the local-form bispectrum, the scale dependence goes as $1/k^2$;
       however, for a modified initial state, this scale dependence can
       become $1/k^3$.
 \item[2.] Anisotropy in the so-called $\mu$-type
	   distortions of the black-body spectrum of the
	   CMB can be correlated with the CMB temperature
	   anisotropy measured on large scales. This correlation can be
	   used to measure the bispectrum in the 
	   squeezed configuration but with a larger value of $k_1/k$
	   than previously thought possible \cite{Pajer:2012vz}. 
\end{itemize}

This paper is organized as follows. In
Section~\ref{sec:reviewing-model}, we review the model under consideration. 
In Section~\ref{sec:calculation}, we give the form of the
bispectrum and comment on potential uncertainties in the results. 
In Section~\ref{sec:approximation}, we
discuss a useful approximation to the bispectrum in the
squeezed configuration. In
Section~\ref{sec:scale-dpdt-bias}, we calculate the signal
of this model in the scale-dependent bias of LSS. In Section
\ref{sec:mu-distortion}, we calculate the signal of this
model in the $\mu$-type distortion of the CMB black-body
spectrum, correlated with the CMB temperature anisotropy on large
scales. Finally, we conclude in Section \ref{sec:discussion}.

Throughout this paper, we shall set $M_{\rm pl}\equiv 1/\sqrt{8\pi
G}\equiv 1$, and use the cosmological parameters given by
the WMAP 5-year best-fit parameters (WMAP+BAO+$H_0$ ML; \cite{Komatsu:2008hk}):
$\Omega_M=0.277$, $\Omega_\Lambda=0.723$, $h=0.702$, $n_s=0.962$, and
$\Delta_\zeta^2(k_0=0.002~{\rm Mpc}^{-1})=2.46\times 10^{-9}$, unless
stated otherwise.

\section{Action and mode function}
\label{sec:reviewing-model}
We consider here single-field slow-roll inflation with a
canonical kinetic term, where the action (to lowest order in slow-roll) can be written as \cite{Malda}
\begin{align}
 \label{eq:acn}
 S = & S_2 + S_3\,,\cr
 S_2 = & \frac{1}{2} \int d^4x \frac{\dot \phi^2}{H^2} 
  [a^3 \dot \zeta^2 - a (\partial \zeta)^2],\cr
  S_3 = & \int d^4x \frac{\dot \phi^4}{H^4} a^5 H \dot \zeta^2 \partial^{-2} \dot \zeta.
\end{align}
We expand the curvature perturbation into creation,
$a^\dagger_{\mbf k}$, and annihilation, $a_{\mbf k}$, operators
(not to be confused with the Robertson-Walker scale factor, $a(t)$):
\begin{equation}
 \zeta({\mbf x},t) = 
\int\frac{d^3{\mbf k}}{(2\pi)^3}\left[a_{\mbf k}u_k(t)e^{i{\mbf
				    k}\cdot{\mbf x}}
+a^\dagger_{\mbf k}u^*_k(t)e^{-i{\mbf
				    k}\cdot{\mbf x}}
\right].
\end{equation}
Usually, one chooses an 
initial state so that a comoving observer in the approximately de Sitter
spacetime observes no particles (i.e., for this observer $a_{\mbf
k}|0\rangle=0$). This implies a positive-frequency mode function given
by
\begin{align}\label{eq:BD_mode_fcn}
  u_k(\eta)
   =  \frac{H^2}{\dot \phi} \frac{1}{\sqrt{2 k^3}} (1+ik \eta) e^{-ik \eta},
\end{align}
where $\eta\equiv \int^t \frac{dt'}{a(t')}$ is the conformal
time; for future reference, we note
\begin{align}\label{eq:BD_mode_fcn_deriv}
  u_k'\equiv\frac{\partial u_k}{\partial \eta} = \frac{H^2}{\dot\phi} \sqrt{\frac{k}{2}} \eta e^{-ik\eta}.
\end{align}

While this is certainly a reasonable assumption, it is an
assumption, and all assumptions must be tested by observations. Thus, a responsible scientist should ask: ``{\it If the initial state of $\zeta$ was 
not in this preferred vacuum state (known as the Bunch-Davies state),
what are the implications for observations?''} Our goal in this paper is
not to construct candidate models of a modified initial state, but to
study phenomenological consequences of such a modification, i.e., to let our
observations tell us about the initial state of quantum fluctuations.

Once we adopt this approach, the next question is: ``How
should we parametrize a modified initial state?'' We will represent a
modified initial state as a Bogoliubov transformation of the above
Bunch-Davies mode function: 
\begin{align}
  \tilde{u}_k(\eta) = \alpha_k u_k(\eta) + \beta_k u^*_k(\eta).
\label{eq:Bogoliubov}
\end{align}
This is not the most general form one can write down (see, e.g.,
\cite{Kundu:Iftn-General-Init}), but it provides us with a reasonable
starting point. In line with our previous goal, we will take the
Bogoliubov coefficients as given rather than trying to derive them from
a fundamental theory.  From the commutation relation of creation and
annihilation operators, the coefficients $\alpha_k$ and $\beta_k$
must satisfy $\lvert\alpha_k\rvert^2 - \lvert\beta_k\rvert^2=1$. 
We also find that the occupation number of particles
$N_k$, i.e., the expected number density of particles with momentum $k$,
is given by $\lvert \beta_k \rvert^2$. 

These Bogoliubov coefficients, $\alpha_k$ and $\beta_k$, encode
information about physics on scales where we have limited information;
thus, they can vary widely without inconsistency. However, we can place
some constraints on the coefficients by demanding that the theory
reproduce the observed power spectrum (including the spectral tilt,
$n_s=0.96\pm 0.01$) and that the energy in the fluctuations not back-react on the background inflaton dynamics \cite{Anderson-Molina-Mottola,Boyanovsky:2006qi,Kundu:Iftn-General-Init}. These requirements can be satisfied in a fairly natural way if we suppose that the coefficients are such that $\langle N_k \rangle \approx N_{k,0} e^{-k^2/k_\text{cut}^2}$, where the cutoff momentum $k_\text{cut}$ must be specified. The values allowed for $N_{k,0}$ depend on the value of $k_\text{cut}$ \cite{Boyanovsky:2006qi}; for $k_\text{cut}\approx\sqrt{\mpl H}$, i.e., the scale of inflation, $N_{k,0}$ can be of order unity. Additionally, if we suppose that the smallest primordial scales observable today come from momenta sufficiently smaller than $k_\text{cut}$, then $\langle N_k \rangle \approx N_{k,0}\equiv N$, i.e., roughly constant in $k$. Remembering that $\langle N_k \rangle = \lvert\beta_k\rvert^2$ and that only the relative phase between $\alpha_k$ and $\beta_k$ is significant, we parametrize
\begin{align}\label{eq:parm_al_bet}
  \alpha_k&\equiv\sqrt{1+N}e^{i\theta_k}, & \beta_k&\equiv\sqrt{N}\,.
\end{align}

There is still uncertainty with respect to $\theta_k$. As explained
 further in \cite{Ganc:2011dy}, we identify two scenarios as plausible
 behaviors: 1) $\theta_k \approx k \eta_0$, where $k |\eta_0|\gg 1$ for
 relevant $k$, and 2)
 $\theta_k\approx\text{const}\equiv\theta$. In the latter scenario, one
 can tune the value of $\theta$ to give larger effects; 
 we will generally show results that assume the value of
 $\theta$ that gives the largest signal. In this sense (and for another
 reason discussed in \cite{Ganc:2011dy}), we consider the former
 scenario to be more conservative. 

\section{Power spectrum and bispectrum}
\label{sec:calculation}
The power spectrum of $\zeta$ on super-horizon scales,
$k\eta\ll 1$, which seeds the observed fluctuations, is given simply by
$P_\zeta(k)=|\tilde{u}_k(\eta\to 0)|^2$ \cite{Ganc:2011dy}, i.e.,
 \begin{align}\label{eq:nbd_pow_spec}
  P_\zeta(k) = \frac{H^4}{\dot \phi^2} \frac{1}{2 k^3} \left| \alpha_k + \beta_k \right|^2,
\end{align}
which becomes (using Eq.~(\ref{eq:parm_al_bet}))
\begin{align}\label{eq:pow-spr-w-phase-ang}
  P_\zeta(k) = \frac{H^4}{\dot \phi^2} \frac{1}{2 k^3} 
      \left(1 + 2 N + 2 \sqrt{N (N + 1)} \cos \theta_k\right).
\end{align}

The calculation of the bispectrum requires more thought. Formally, it is
given by \cite{Malda} 
\begin{eqnarray}\label{eq:in-in-formalism}
\nonumber
& & \langle \zeta_{\mbf{k}_1}(t)\zeta_{\mbf{k}_2}(t)\zeta_{\mbf{k}}(t)\rangle\\
&=&-i\int_{t_0}^tdt'\langle[\zeta_{\mbf{k}_1}(t)\zeta_{\mbf{k}_2}(t)\zeta_{\mbf{k}}(t),H_{\rm int}(t')]\rangle,
\end{eqnarray}
where the interaction Hamiltonian, $H_{\rm int}$, is given
by $\int dt' H_{\rm int}(t')=-S_3$ with $S_3$ given by
Eq.~(\ref{eq:acn}). We would then specify the initial state at the
initial time, $t_0$, or equivalently at the initial conformal time, $\eta_0$.
For the action given by Eq.~(\ref{eq:acn}), one finds 
\begin{eqnarray}
\nonumber
 B_\zeta(k_1,k_2,k)&=&2i\frac{\dot{\phi}^4}{H^6}
\frac{k_1^2k_2^2+k_2^2k^2+k^2k_1^2}{k_1^2k_2^2k^2}\tilde{u}_{k_1}\tilde{u}_{k_2}\tilde{u}_{k}\\
& &\times\int_{\eta_0}^\eta
 \frac{d\eta'}{(\eta')^3}\tilde{u}'^*_{k_1}\tilde{u}'^*_{k_2}\tilde{u}'^*_{k} + \operatorname{c.c.}\,.
\label{eq:integral}
\end{eqnarray}
In this paper, dots will denote derivatives with respect to $t$ and
primes will denote derivatives with respect to $\eta$.
For the standard calculation, we take the Bunch-Davies initial vacuum
state, given by $\alpha_k=1$ and $\beta_k=0$, for all modes into the
infinite past, $\eta_0\to - \infty$ (i.e., $t\to 0$). For this case,
there is an accepted prescription for calculations: we take $\eta_0\to
\eta_0+i\epsilon|\eta_0|$, giving $\eta$ an imaginary component when its
absolute value is large \cite{Malda}. The exponential terms in the
integrand like $e^{i(k_1+k_2+k)\eta_0}$ (see
(\ref{eq:BD_mode_fcn_deriv}) for their origin) would ordinarily
oscillate rapidly at very early times but are suppressed by the
imaginary part of $\eta_0$. Note that this suppression depends on $k_1 +
k_2 + k > 0$. 

However, when we allow for a more general initial state, we can have $\beta_k\neq 0$ resulting in terms like $e^{i(-k_1+k_2+k)\eta_0}$, $e^{i(-k_1-k_2+k)\eta_0}$, etc. Furthermore, one may object (e.g., for reasons of renormalizability) to setting initial conditions in the infinite past, especially if some of the modes are excited (i.e., $\beta_k\neq 0$); instead, one might prefer that initial conditions be set at some finite time. If we ignore this objection for a moment, one can still suppose that $\eta_0\to \eta_0+i\epsilon|\eta_0|$. By triangle inequalities (e.g., $k_1 \leq k_2 + k$, etc.), the exponentials are still suppressed except at the precise folded limit $k_1 = k_2 + k$ (note that this would result in Eq.~(\ref{eq:nbd-bisp}) but without the exponentials).

In this paper, however, we will generally take the objection seriously and suppose that initial conditions were not set infinitely far in the past. Unfortunately, this draws us into an area of active research which does not offer a definite formalism for calculations. Here, as in \cite{Holman:2007na} (though see \cite{Meerburg:2009ys,Meerburg:2009fi,*Ashoorioon:2010xg}), we will adopt the ``Boundary Effective Field Theory'' approach to non-Bunch Davies initial
conditions \cite{Schalm:2004qk,*Greene-BEFT}, which like the other
available approaches is not without problems (e.g.,
\cite{Easther:BEFT-v-NPH}). In this approach, one cuts off the integral
given in Eq.~(\ref{eq:in-in-formalism}) at a finite $\eta_0$, where the initial conditions are set.

We shall assume that for excited modes $k$ (i.e., where $\beta_k\neq
0$), $k |\eta_0|\gg 1$ so that $k$ was deep inside the horizon at the
initial time. This can be explained as expressing the requirement that
the process for mode excitation was causal and thus, could only excite
subhorizon modes. 

Performing the integral given in Eq.~(\ref{eq:integral}), one obtains the bispectrum \cite{Ganc:2011dy}
\begin{align}\label{eq:nbd-bisp}
 &B_\zeta(k_1,k_2,k)=\frac{1}{2} \frac{H^6}{\dot \phi^2} 
\frac{k_1^2k_2^2+k_2^2k^2+k^2k_1^2}{k_1^3k_2^3k^3}
\cr
  & \times\Re\Big[\frac{1}{k_1 + k_2 + k} 
     F_{\alpha\alpha\alpha}
       \left(1 - e^{i (k_1 + k_2 + k) \eta_0}\right) 
\cr
& \phantom{\times\Re}  + \frac{1}{k_1 + k_2 - k} 
     F_{\alpha\alpha\beta}
       \left(1 - e^{i (k_1 + k_2 - k) \eta_0}\right)
\cr
  &\phantom{\times\Re}
   + \frac{1}{k_1 - k_2 + k} 
     F_{\alpha\beta\alpha}
       \left(1 - e^{i (k_1 - k_2 + k) \eta_0}\right)
\cr
  &\phantom{\times\Re}
   + \frac{1}{- k_1 + k_2 + k} 
     F_{\beta\alpha\alpha}
       \left(1 - e^{i (- k_1 + k_2 + k) \eta_0}\right) \Big],
\end{align}
where
\begin{align}
  F_{XYZ} \equiv (\alpha_{k_1} + \beta_{k_1}) (\alpha_{k_2} + \beta_{k_2}) (\alpha_{k} + \beta_{k}) 
     X_{k_1}^* Y_{k_2}^* Z_{k}^* 
\cr
   - (\alpha_{k_1}^* + \beta_{k_1}^*) (\alpha_{k_2}^* + \beta_{k_2}^*) (\alpha_{k}^* + \beta_{k}^*) 
     X^\text{C}_{k_1} Y^\text{C}_{k_2} Z^\text{C}_{k},
\end{align}
for $\alpha^\text{C} \equiv \beta$, $\beta^\text{C} \equiv \alpha$; $F_{XYZ}$ gives information about the initial conditions at $\eta_0$. Note
that we ignore a field redefinition term (derived in \cite{Malda}) that
is negligibly small for the purposes of this paper. 

First, note that we recover the standard Bunch-Davies result
\cite{Malda} if we set $F_{\alpha\alpha\alpha}=1$,
$F_{\alpha\alpha\beta}=F_{\alpha\beta\alpha}=F_{\beta\alpha\alpha}=0$
, and $\eta_0\to -(1-i\epsilon)\infty$. In the squeezed limit, $k\ll k_1\approx k_2$ and we get 
$B_\zeta\to 2 \frac{\dot{\phi}^2}{2H^2}P_\zeta(k_1)P_\zeta(k)$, where 
$\frac{\dot{\phi}^2}{2H^2}\approx 10^{-2}$ is the slow-roll
parameter, which is equivalent to  $f_{\rm NL}={\cal
O}(10^{-2})$
(see Eq.~(\ref{eq:locallimit})). If we restore the field-redefinition piece we ignored, we obtain  the full standard squeezed-limit
bispectrum: $B_\zeta\to (1-n_s)P_\zeta(k_1)P_\zeta(k)$ \cite{Malda}.

Since we have assumed $k |\eta_0|\gg 1$, the exponentials in the bispectrum (\ref{eq:nbd-bisp}) oscillate rapidly and can, to a decent approximation, be ignored. Then, one sees that the bispectrum peaks in the so-called ``folded triangle configuration,'' where one of the
wavenumbers is approximately equal to the sum of the other two, i.e., 
$k\approx k_1+k_2$, $k_2\approx k_1+k$, or $k_1\approx k_2+k$; this was noted earlier by
\cite{Chen:2006nt,Holman:2007na,Meerburg:2009ys}. Since the local bispectrum has no corresponding peak, this regime provides a way to distinguish the shape of this bispectrum from a purely local form. We shall come back to this point in Section~\ref{sec:discussion}.

We can also investigate the squeezed configuration $k\ll k_1\approx k_2$; this configuration is in fact a special case of the folded limit $k_1\approx
k_2+k$ when we additionally suppose that $k$ is much smaller than $k_1$ or $k_2$. In this limit, the third and fourth terms are larger than the first and second by a factor of $k_1/k\gg 1$; the bispectrum becomes $B_\zeta\propto 8
\frac{\dot{\phi}^2}{2H^2}\frac{k_1}{k}P_\zeta(k_1)P_\zeta(k)$ (with a proportionality factor $|\alpha_{k_1} + \beta_{k_1}|^{-2} |\alpha_{k} + \beta_{k}|^{-2}\Re\left[F_{\alpha\beta\alpha} (1-e^{i k \eta_0})\right]$). Note that this is enhanced relative to the local form in the squeezed configuration \cite{Agullo:2010ws}.

We should highight that the exponential terms cannot be completely ignored \cite{Meerburg:2009ys} because they prevent the bispectrum from blowing up in the folded limit. In particular, the factor $\frac{1}{-k_1+k_2+k}[1-e^{i(-k_1+k_2+k)\eta_0}]$, which seems to blow up in the folded limit if one ignores the exponential, actually goes as
$-i\eta_0 + \mathcal O ((-k_1+k_2+k) \eta_0^2)$. Accounting for this
behavior plays a role in the usefullness of the approximation we
demonstrate in the next section. 

\section{Approximation to the bispectrum in the squeezed configuration}
\label{sec:approximation}

While the full form of the bispectrum given by
Eq.~(\ref{eq:nbd-bisp}) is complicated, the observables that we shall
discuss in this paper (the scale-dependent halo bias in LSS and the anisotropy
in the $\mu$-type distortion of the CMB black-body spectrum) depend primarily
on the squeezed configuration, $k\ll k_1\approx k_2$. Therefore, it is useful
to find an accurate approximation to the bispectrum in the squeezed
configuration.

In \cite{Ganc:2011dy}, the author expanded to the lowest order in $k$
($k$ here is equal to $k_3$ in \cite{Ganc:2011dy}) after averaging over
the exponential. Specifically, he approximated
\begin{align}\label{eq:orig-approx}
  \frac{1-e^{i(k_1 -k_2 +k) \eta_0}}{k_1 -k_2 +k}\approx \frac{1}{k}\,.
\end{align}
This result is also consistent with a prescription of
ignoring oscillating terms by taking $\eta_0\to \eta_0+i\epsilon|\eta_0|$ for large $|\eta_0|$, as discussed in the previous section.

When we do not ignore oscillating terms, the approximation
demonstrates the correct scaling on large scales but it is off by a
factor. This arises because the approximation does not properly account
for the oscillatory behavior of Eq.~\eqref{eq:orig-approx} at small
$k$. 

Fortunately, we can come up with a better approximation. Observe that,
when calculating observables, the bispectrum is usually
multiplied by a 
function and then integrated over some of the wavenumbers (see, e.g.,
Eq.~\eqref{eq:bias-formula} below). Let us focus on the integral over $k_2$.
Note that the limits of integration for $k_2$ are $k_2\in
[|k_1-k|,k_1+k]$; in the squeezed limit, the function multiplying the
bispectrum will vary little over this small range, while the oscillatory
terms like the left hand side of Eq.~\eqref{eq:orig-approx} will vary very
rapidly. Thus, we can perform the $k_2$ integral only over the rapidly oscillating term, e.g.,
\begin{align}
  \int d&k_1 \int_{k_1-k}^{k_1+k} dk_2\, \cdots \frac{1-e^{i(k_1 -k_2 +k) \eta_0}}{k_1 -k_2 +k} \approx\cr{}
  &{}\approx\int dk_1 \, \cdots\Big|_{k_2= k_1} \left(\int_{k_1-k}^{k_1+k} dk_2 \frac{1-e^{i(k_1 -k_2 +k) \eta_0}}{k_1 -k_2 +k} \right)\,.\cr
\end{align}
For this integral, we find
\begin{align*}
  \int_{k_1-k}^{k_1+k} &dk_2 \frac{1-e^{i(k_1-k_2+k)\eta_0} }{k_1-k_2+k}
      \cr
   &= \left[\gamma - \operatorname{Ci}(-2 k \eta_0)+\log(-2 k \eta_0)\right]+ i \operatorname{Si}(-2 k \eta_0)\cr
   &\approx \gamma - \operatorname{Ci}(-2 k \eta_0)+\log(-2 k \eta_0)\,,
\end{align*}
where $\gamma\approx 0.5772$ is Euler's constant,
$\operatorname{Ci}(z)\equiv-\int_z^\infty dt\cos(t)/t$ is the cosine
integral, and $\operatorname{Si}(z)\equiv\int_0^z dt\sin(t)/t$ is the sin integral (which is $\sim\pi/2$ for $z>1$); in the last line, we have dropped the second term since it becomes increasingly unimportant for large $k\eta_0$.

If we perform this new approximation, we find that, for
$\theta\approx k\eta_0$, the chief contributor to the squeezed bispectrum looks like 
\begin{align}\label{eq:n_not_const_approx}
  B_{\zeta,k\ll k_1}^{\theta_k\approx k \eta_0}(k_1, k_2, k)
  &\approx \frac{H^6}{\dot \phi^2} \frac1{k_1 k_2
 k^4}\cr
&\phantom{\approx}\times N(1+N)\cr
&\phantom{\approx}\times \frac12
    [\gamma-\operatorname{Ci}(-2 k \eta_0) +\log(-2 k \eta_0)].\cr
\end{align}
For $\theta_k\approx\text{const}\equiv\theta$, we find
\begin{align}\label{eq:n_const_approx}
  B_{\zeta,k\ll k_1}^{\theta_k\approx\text{const}}(k_1, k_2, k)
    \approx &\frac{H^6}{\dot \phi^2} \frac{1}{k_1 k_2 k^4}\cr
   &\times  \Big[N (1 + N) (3 - \cos 2 \theta) \cr
   &\phantom{\times \Big[}
       + \sqrt{N (1 + N)} (1 + 2 N) \cos \theta\Big]\cr
   &\times \frac12
      [\gamma - \operatorname{Ci}(-2 k \eta_0)+\log(-2 k \eta_0)].\cr
\end{align}
These two equations provide useful approximations to the bispectrum from
a modified initial state in the squeezed configuration. 

Note that we can also view this approximation as finding a sort of
average for the left hand side of (\ref{eq:orig-approx}), i.e., that
\begin{align*}
   &\left[\frac{1-e^{i(k_1-k_2+k)\eta_0} }{k_1-k_2+k}\right]_{avg}\cr
   &\hspace{4em}\equiv \frac{1}{2k} \int_{k_1-k}^{k_1+k} dk_2 \frac{1-e^{i(k_1-k_2+k)\eta_0} }{k_1-k_2+k}
      \cr
   &\hspace{4em}\approx \frac{\gamma - \operatorname{Ci}(-2 k \eta_0)+\log(-2 k \eta_0)}{2k}\,.
\end{align*}
Thus, ignoring the oscillating terms in the bispectrum (as in \cite{Ganc:2011dy}) is equivalent to neglecting a factor $\frac12[\gamma-{\rm Ci}(-2k\eta_0)+\log(-2k\eta_0)]$, so that the regime investigated in \cite{Ganc:2011dy} differs from the one here by a factor of order unity. 

\section{Scale-dependent bias}
\label{sec:scale-dpdt-bias}
How can we measure $B_\zeta(k_1,k_2,k)$ observationally? 
Obvious observables are the bispectrum of the CMB temperature and
polarization anisotropy, and that of the matter density distribution in
LSS. These observables are (in linear theory) related to
$B_\zeta(k_1,k_2,k)$ in a straightforward way
\cite{Komatsu:2001rj,Sefusatti:2007ih}.

A much less obvious observable is the {\it power spectrum} of dark matter halos (in
which galaxies and clusters of galaxies would be formed). Dark matter
halos are formed only at the locations of peaks of the underlying matter
distribution. While the power spectrum of the underlying matter
distribution is insensitive to the bispectrum, the power spectrum of
{\it peaks} is sensitive to the bispectrum as well as to higher-order
correlation functions \cite{Grinstein:1986en,*Matarrese:1986et}. This
leads to a remarkable prediction: one can use the observed power
spectrum of the distribution of galaxies (and of clusters of galaxies)
to measure  
the bispectrum of primordial fluctuations
\cite{Dalal:2007cu,Slosar:2008hx,Matarrese:2008nc}.

In general, as the power spectrum of peaks (hence halos) is different from
that of the underlying matter distribution, we say that halos are biased
tracers of the underlying matter distribution \cite{Kaiser:1984sw}. The
degree of bias is often parametrized by the so-called ``bias factor,''
$b(k)$, defined as
\begin{equation}
 b^2(k)\equiv \frac{P_{\rm halo}(k)}{P_{\rm matter}(k)}.
\end{equation}
Alternatively, one may define $b(k)$ as the ratio of the matter-halo
cross power spectrum to the matter power spectrum.

On large scales, where the matter density fluctuations are still in the
linear regime, $b(k)$ approaches a constant for Gaussian matter density
fluctuations, $b(k)\to b_1$. However, the presence of the primordial
bispectrum leads to a non-trivial $k$-dependence in $b(k)$, and this is
called a ``scale-dependent bias.'' 

Building on the previous work on the peak statistics
\cite{Grinstein:1986en,*Matarrese:1986et,Matarrese:2008nc}, Desjacques,
Jeong, and Schmidt arrived at the following formula for $b(k)$
\cite{Desjacques:2011mq,Desjacques:2011jb}: 
\begin{eqnarray}
\nonumber
\Delta b(k,R)
= 2\frac{{\cal F}_R(k)}{{\cal
M}_R(k)}\left[(b_1-1)\delta_c+\frac{d\ln{\cal F}_R(k)/d\ln
	 R}{d\ln\sigma_R/d\ln R}\right],\\
\label{eq:bias}
\end{eqnarray}
where $\Delta b(k,R)\equiv b(k,R)-b_1$, $\delta_c=1.686$, $R$ is related
to the mass 
of halos under consideration as $M=\frac{4\pi}3\Omega_M\rho_cR^3$, and
$\rho_c=2.775\times 10^{11}~h^2~M_\odot~{\rm Mpc}^{-3}$ is the
present-day critical density of the universe. The various functions are
defined by 
\begin{eqnarray}
\nonumber
 {\cal F}_R(k)&\equiv&
  \frac1{4\sigma_R^2P_\zeta(k)}\int\frac{d^3k_1}{(2\pi)^3}{\cal
  M}_R(k_1){\cal
  M}_R(|\mathbf{k}_1+\mathbf{k}|)\\
& &\qquad\qquad\quad \times B_\zeta(k_1,|\mathbf{k}_1+\mathbf{k}|,k),\\ 
\sigma_R^2&\equiv& \int\frac{d^3k}{(2\pi)^3}P_\zeta(k){\cal M}^2_R(k),\\
 {\cal M}_R(k)&\equiv& \frac{2 k^2D(z)}{5\Omega_M H_0^2}T(k) W_R(k)\,,\\
 W_R(k)&\equiv& \frac{3 j_1(kR)}{kR},\\
\end{eqnarray}
where $T(k)$ is the linear transfer function normalized such that
$T(k)\to 1$ for $k\to 0$, and $D(z)$ is the growth
factor of linear density fluctuations normalized such that $(1+z)D(z)\to
1$ during the matter era. (For example, $D(0)=0.7646$ for $\Omega_M=0.277$, $\Omega_\Lambda=0.723$, and $w=-1$.)
 
Before we show the numerical calculations, we will first try to analytically explore Eq.~(\ref{eq:bias}) for the case of a modified initial state, allowing us to estimate the bispectrum shape. 

An important observation in what follows is that $W_R(k)$ oscillates
rapidly for $k\gtrsim 1/R$, so that the integral of ${\cal F}_R(k)$ is
dominated by $k_1\approx 1/R$. Therefore, when we are interested in
$k\ll k_1$, the integral of ${\cal F}_R(k)$ is dominated by the squeezed
configuration, and is approximated as
\begin{align}
 {\cal F}_R(k)\approx
  \frac1{4\sigma_R^2P_\zeta(k)}\int\frac{d^3k_1}{(2\pi)^3}&{\cal
  M}_R^2(k_1)B_\zeta(k_1,k_1,k).
\end{align}
We can insert the squeezed configuration local bispectrum and calculate
\begin{equation}
 {\cal F}_R(k)\approx \frac35f_{\rm NL}.
\end{equation}
The second term in the parenthesis in Eq.~(\ref{eq:bias})
vanishes for this case and we find $\Delta b(k,R)\propto 1/k^2$
\cite{Dalal:2007cu,Slosar:2008hx,Matarrese:2008nc}. 

For the bispectrum for a modified initial state, which goes as $1/k_1^2k^4$, we instead find
\begin{equation}
 {\cal F}_R(k)\propto\frac{\bar{k}_{1}(R)}{k},
\end{equation}
where
\begin{equation}
 \bar{k}_1(R)\equiv
  \frac1{\sigma_R^2}\int\frac{d^3k_1}{(2\pi)^3}k_1{\cal
  M}^2_R(k_1)P_\zeta(k_1).
\label{eq:k1}
\end{equation}
One may interpret $\bar{k}_1$ as a characteristic wavenumber for the
short-wavelength mode in the squeezed configuration. Thus, we expect the modified-state bispectrum to produce  a scale-dependent bias which grows faster (by a factor
of  $\bar{k}_1/k\approx 1/(kR)$) for small  values of $k$ than that for the
local-form bispectrum.  

What about the second term in the brackets in Eq.~(\ref{eq:bias})? If we
note that the extra $k$ factor in the integrand of Eq.~(\ref{eq:k1}) (as
compared with the integrand for $\sigma_R^2$) is evaluated at roughly
$1/R$, we get $\bar k_1(R)\approx 1/R$ and $d\ln {\cal F}_R(k)/d\ln
R\approx -1$.

On the other hand, $\sigma_R^2$ is dominated by the power spectrum of
 matter density fluctuations at $k\approx 1/R$. Approximating the power
 spectrum of  matter density fluctuations as a power-law near $k\approx
 1/R$, i.e., ${\cal M}_R^2(k)P_\zeta(k)|_{k\approx R^{-1}}\propto k^{n_{\rm
 eff}(R)}$, one 
 obtains $d\ln\sigma_R/d\ln R=-[n_{\rm eff}(R)+3]/2$. For example,
 $n_{\rm eff}(R)=-2.2$, $-1.8$, and $-1.6$ for $R=1$, 5, and
 10~$h^{-1}$~Mpc (or $M=3.2\times 10^{11}$, $4.0\times 10^{13}$, and
 $3.2\times 10^{14}~h^{-1}~M_\odot$), respectively.

Therefore, while this second term
changes the amplitude of $\Delta b(k,R)$ by a factor of $1+\frac{2}{[n_{\rm
eff}(R)+3]\delta_c(b_1-1)}$, it does not change 
the $k$-dependence of $\Delta b(k,R)$. We thus expect the
 $k$-dependence of the scale-dependent bias for a modified initial state
 to be given by $\Delta b(k,R)\propto 1/k^3$. This scaling was also
predicted by \cite{Chialva:2011hc}. 

In principle, the second term can change the
amplitude of $\Delta b(k,R)$ by a large factor for low-mass halos whose bias
is closer to unity
\cite{Desjacques:2011mq,*Desjacques:2011jb}. Nevertheless, as we are
focused on the shape of  
$\Delta b(k,R)$ rather than on the amplitude, we will ignore this factor.
Then, Eq.~(\ref{eq:bias}) simplifies to 
\begin{align}\label{eq:bias-formula}
  \frac{\Delta b(k,R)}{b_1-1}=& 
\frac{1}{8\pi^2\sigma_R^2}\frac{\delta_c}{\mathcal{M}_R(k)P_\zeta(k)k} 
    \int_0^\infty \mathrm{d}k_1 \,k_1 \mathcal{M}_R(k_1)\cr
&\times \int_{\lvert k_1 - k\rvert}^{k_1+k}\mathrm{d} k_2 \, k_2
      \mathcal{M}_R(k_2) 
     B_\zeta(k_1,k_2,k)\cr
=& \frac{1}{20\pi^2D(z)\tilde{\sigma}_R^2}\frac{\delta_c}{\Omega_MH_0^2k^3P_\zeta(k)T(k)W_R(k)} \cr
&\times    \int_0^\infty \mathrm{d}k_1\, k_1^3 T(k_1)W_R(k_1)\cr
&\times \int_{\lvert k_1 - k\rvert}^{k_1+k}\mathrm{d} k_2 \,k_2^3
      T(k_2)W_R(k_2)
     B_\zeta(k_1,k_2,k),
\end{align}%
which agrees with the formula first derived  by
\cite{Matarrese:2008nc}. Here, $\tilde{\sigma}_R\equiv \sigma_R/D(z)$,
which is independent of $z$.

\begin{figure}[t]
  \centering
  \scalebox{0.5} {\includegraphics{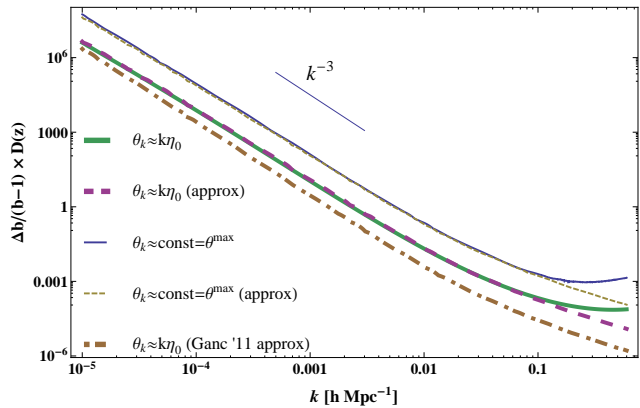}}
  \caption{Scale-dependent halo-bias from single-field inflation with a
 non-standard initial state, using a smoothing scale of
 $R=1~h^{-1}$~Mpc. The occupation number is $N=0.5$, the slow-roll 
 parameter $\epsilon=0.01$, and the initial conformal time
 $|\eta_0|=1.0\times10^6\ \text{Mpc}$ (the results are insensitive to the exact
 choice of $\eta_0$, so long as it is large). The bottom three (thicker) lines
 show the more natural case 
 where $\theta_k\approx k \eta_0$, while the top two (thinner) lines
 show the case 
 when $\theta_k\approx\text{const}$ is chosen to give the maximal halo
 bias. The dashed lines show the new approximations given by
 Eqs.~\eqref{eq:n_not_const_approx} and \eqref{eq:n_const_approx},
 while the dot-dashed line shows the approximation used in
 \cite{Ganc:2011dy} (which is equal to Eqs.~\eqref{eq:n_not_const_approx} and
 \eqref{eq:n_const_approx} without the last lines).} 
  \label{fig:thet-not-const_v_thet-const}
\end{figure}

\begin{figure}[t]
  \centering
  \scalebox{0.5}{\includegraphics{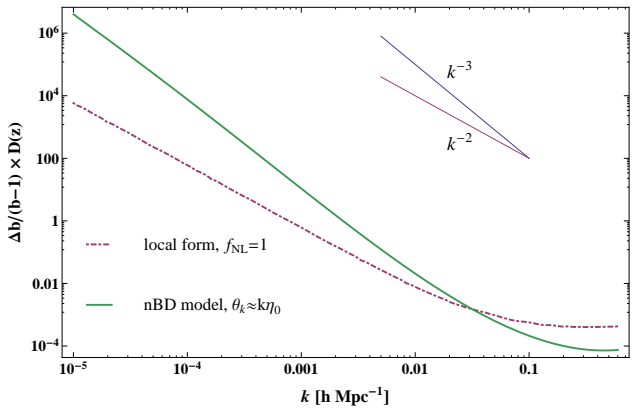}}
  \caption{Scale-dependent bias from the local-form bispectrum (dot-dashed
 line) versus the modified initial-state case described herein (solid
 line). The parameters here are the same as in Fig.
 \ref{fig:thet-not-const_v_thet-const}, with $f_{\text{NL}}=1$ for the
 local-form bispectrum. The difference in scaling between the models is quite evident.}
  \label{fig:loc-v-nBD}
\end{figure}

To evaluate Eq.~(\ref{eq:bias-formula}) we will use $R=1~h^{-1}$~Mpc
(corresponding to $M=3.2\times 10^{11}~h^{-1}~M_\odot$). Also, in order
to determine the factor $H^6/\dot\phi^2$ which appears in the
bispectrum, we use the WMAP 5-year normalization,
$k^3_0P_\zeta(k_0)/(2\pi^2)=2.41\times 10^{-9}$ for $k_0=0.002~{\rm
Mpc}^{-1}$ \cite{Komatsu:2008hk}, in Eq.~\eqref{eq:pow-spr-w-phase-ang}:
\begin{equation}
 \frac{k_0^3P_\zeta(k_0)}{2\pi^2} =
  \frac1{8\pi^2}\frac{H^2}{\epsilon}\left(1+2N+2\sqrt{N(N+1)}\cos\theta\right),
\label{eq:normalization}
\end{equation}
where $\epsilon\equiv(1/2)\dot\phi^2/H^2$ is the slow-roll
parameter; for $\theta_k\approx k\eta_0$, the term in parenthesis simply
becomes $(1+2N)$. This relation gives $H^2$
for a given $\epsilon$, $N$, and $\theta$.

We now insert the full bispectrum (Eq.~\eqref{eq:nbd-bisp})
into Eq.~\eqref{eq:bias-formula} and numerically integrate for the halo
bias. Fig.~\ref{fig:thet-not-const_v_thet-const} 
shows the results of the numeric integration for $\epsilon=0.01$,
$N=0.5$, and $\theta_k$ for both scenarios ($\theta_k=k\eta_0$ and
$\theta=\mbox{const}$, with $\theta$ chosen to maximize $\Delta b$).  We do find the expected $k^{-3}$ scaling, which can also be seen by comparison with the local form in Fig.~\ref{fig:loc-v-nBD}. 

Fig.~\ref{fig:thet-not-const_v_thet-const} also shows  the halo bias as
calculated from the approximation given in
Eq.~\eqref{eq:n_not_const_approx}, as well as from the earlier
approximation from \cite{Ganc:2011dy} (which is
Eq.~\eqref{eq:n_not_const_approx} without the last line). One sees that
the full calculation and the new approximation are very similar except
on very small scales  (near the smoothing scale); by contrast, the old
approximation is off in absolute scale. (There is also a slight change
in shape, due to the terms that depend on $k\eta_0$).

\section{$\mu$-type distortion of the black-body spectrum of CMB}
\label{sec:mu-distortion}
\subsection{Motivation and background}

Diffusion damping of acoustic waves heats CMB photons and creates spectral distortions in the black-body spectrum of the CMB
\cite{Sunyaev:1970er,*1970ApSS...9..368S,*1991ApJ...371...14D}. However, this distortion is erased,
maintaing a black-body spectrum for the CMB, as long as photon
non-conserving processes are effective. According to \cite{Hu:1992dc},
double-Compton 
scattering ($e^- + \gamma \to e^- + 2\gamma$) is an effective 
thermalization process for $z> z_i\approx 2\times 10^6$. After this epoch,
however,  this process shuts off and the spectral distortions from
diffusion damping cannot be smoothed from the CMB spectrum. Since
elastic Compton scattering ($e^- + \gamma \to e^- + 
\gamma$) continues to be effective until $z_f\approx 5\times10^4$, 
the photons can still achieve equilibrium but with a conserved photon
number. The result is a Bose-Einstein
distribution with a non-zero chemical potential, $\mu$ (rescaled by $k_BT$ to be dimensionless), an effect known as the ``$\mu$-type distortion'' of the black-body spectrum of the CMB, and it affects the distribution by
\begin{equation}
 \frac1{e^{h\nu/(k_BT)}-1}\to \frac1{e^{h\nu/(k_BT)+\mu}-1}\,;
\end{equation}
a positive $\mu$ reduces the number of photons at low frequencies.
Finally, after $z_f$, even elastic Compton scattering is inefficient and
photons fall out of kinetic equilibrium with electrons, leaving only the
so-called ``$y$-type distortion'' \cite{1969Ap&SS...4..301Z,*Sunyaev:1972eq}. As it
would be difficult to distinguish among the $y$-distortions created by
the heating of CMB photons due to diffusion damping, by the cosmic
reionization ($z\approx 10$), and by the thermal Sunyaev-Zel'dovich effect
\cite{1969Ap&SS...4..301Z,*Sunyaev:1972eq} from groups and clusters of
galaxies 
($z\lesssim 3$) \cite{Refregier:2000xz}, we shall focus on the $\mu$-type
distortion in this paper.

Diffusion damping occurs near the damping scale given as follows. Over
the redshifts 
of interest, $z\approx 5\times10^4-2\times 10^6$, the expansion rate of
the universe is dominated by radiation, $H(z)\propto (1+z)^2$, and the
effect of baryon density on the photon-baryon fluid is
negligible. Therefore, the damping scale, $k_D$, is given by 
\cite{Silk:1967kq,*1983MNRAS.202.1169K,*weinberg:COS}
\begin{align}
k_D^{-2}&=\int_0^\eta d\eta'  \frac{8}{45\sigma_Tn_ea}
&=-\int_\infty^z dz'  \frac{8(1+z)}{45\sigma_Tn_eH},
\end{align}
which gives 
\begin{equation}
 k_D\approx 130~[(1+z)/10^5]^{3/2}~{\rm Mpc}^{-1}.
\end{equation}
Meanwhile, the heat generated by diffusion damping, $Q$, is given by
\begin{equation}
 Q=\frac14\rho_\gamma\langle\delta^2_{\gamma}\rangle,
\end{equation}
where $\rho_\gamma$ is the photon energy density and $\delta_\gamma$ is
the photon energy density contrast. The coefficient $1/4$ merits further explanation. Naively, it would be $c_s^2 = 1/3$, where  $c_s=1/\sqrt{3}$ is the sound speed of the photon
fluid. However, a recent computation using second-order perturbation theory \cite{Chluba:2012gq} reveals that we need an additional factor of $3/4$, yielding the number above. This heat is then
converted into $\mu$ as 
\begin{equation}
 \mu\approx 1.4\int_{z_f}^\infty
  dz\frac1{\rho_\gamma}\frac{dQ}{dz}e^{-(z/z_i)^{5/2}}\approx 
\frac{1.4}{4}\left[\langle\delta^2_{\gamma}\rangle(z_i)-\langle\delta^2_{\gamma}\rangle(z_f)\right]. 
\end{equation}
The diffusion damping scales at $z_i$ and $z_f$ are given by
$k_D(z_i)\approx 12000~{\rm Mpc}^{-1}$ and $k_D(z_f)\approx
46~{\rm Mpc}^{-1}$, 
respectively. Therefore, the $\mu$-type distortion is created by the
(squared) photon density perturbation on very small scales. This property
allows us to probe the power spectrum on such small scales \cite{Sunyaev:1970er,*1970ApSS...9..368S,*1991ApJ...371...14D,Hu:1994bz,*Chluba:2011hw,*Khatri:2011aj,*Chluba:2012we}.

Pajer and Zaldarriaga recently pointed out that a non-zero bispectrum in
the squeezed configuration makes the distribution of $\mu$ on the sky
{\it anisotropic}, and that this anisotropy of $\mu$ is correlated with the
temperature anisotropy of the CMB, which measurements are on scales much
larger than the damping scale \cite{Pajer:2012vz}. This allows us to
measure the bispectrum in the 
squeezed configuration with a larger value of $k_1/k$ than previously
thought possible. The smallest possible wavenumber one can measure from
the CMB anisotropy in the sky corresponds to the quadrupole, i.e.,
$k\approx 2/r_L\approx 1.4\times 10^{-4}~{\rm Mpc}^{-1}$, where
$r_L\approx 14000$~Mpc is the comoving distance to the last scattering
surface at $z_L=1090$. This 
gives $k_1/k\approx k_D/k=3.3\times 10^5-8.6\times 10^7$, which is far greater
than that accessible from the temperature anisotropy of the CMB in $l=2-3000$,
i.e., $k_1/k = 1 - 1500$, or that accessible from the scale-dependent
bias of LSS: $k_1/k\approx 1/(kR)\approx 
10^3$ for $k\approx 10^{-3}~h~{\rm Mpc}^{-1}$ (the lowest wavenumber
that can be plausibly measured from the LSS data in near future) and
$R\approx 1~h~{\rm Mpc}$. 

\subsection{Cross-power spectrum of CMB temperature anisotropy and
  $\mu$-type distortion}
First, we decompose the CMB temperature anisotropy on the sky into
spherical harmonics: $\delta
T(\hat{n})/T=\sum_{lm}a^T_{lm}Y_{lm}(\hat{n})$. The
spherical harmonics coefficients are related to the primordial curvature
perturbation as
\begin{align}
  a^{T}_{lm} = \frac{12\pi}{5}(-i)^{l}\int
 \frac{d^{3}k}{(2\pi)^{3}}\zeta(\mbf k)g_{Tl}(k)Y^{\ast}_{lm}(\hat k)\,,
\label{eq:almT}
\end{align}
where $g_{Tl}(k)$ is the radiation transfer function. Our sign
and normalization are such that $g_{Tl}(k)\to -j_l(kr_L)/3$ in the
Sachs-Wolfe limit. In other words, $\delta T(\hat{n})/T\to
-\zeta(\hat{n}r_L)/5$ in the Sachs-Wolfe limit. However, we will not use
the Sachs-Wolfe limit (except for comparison), and instead use
$g_{Tl}(k)$ as computed from a linear 
Boltzmann code \footnote{A code for
calculating the radiation transfer function is available at {\sf
http://www.mpa-garching.mpg.de/\textasciitilde{}komatsu/CRL/}. This code
is based on CMBFAST \cite{Seljak:1996is}}.

Next, we similarly decompose the distribution of $\mu$ measured on the
sky, $\mu(\hat{n})$, into spherical harmonics:
$\mu(\hat{n})=\sum_{lm}a^{\mu}_{lm}Y_{lm}(\hat{n})$. Following
\cite{Pajer:2012vz}, we write
\begin{align}
  a^{\mu}_{lm}
   =&18\pi (-i)^{l}\int \frac{d^{3}k_{1}d^{3}k_2}{(2\pi)^{6}} 
    Y^{\ast}_{lm}(\hat k) \zeta(\mbf k_{1}) \zeta(\mbf k_{2}) 
     W \left(\frac{k}{k_s}\right) \times \cr
   &\times j_{l}(k r_L) \langle\cos \left(k_1 r\right) \cos \left(k_2 r\right)\rangle_{p}
    \left[e^{-(k_{1}^{2}+k_{2}^{2})/k_{D}^{2}(z)}\right]^{z_i}_{z_f}\,,
\label{eq:almM}
\end{align}
with $\mbf k_1 + \mbf k_2 + \mbf k = 0$. (Note that the
coefficient of 
our expression is $18\pi$ instead of $24\pi$ because of the factor of
$3/4$, mentioned earlier, from \cite{Chluba:2012gq}). Here, $W(x)\equiv
3j_1(x)/x$ is a filter function; $k_s$ is the scale over which the
damped acoustic waves are averaged to give heat (and which we will take
to be equal to $k_{D}(z_f)$ to obtain a lower bound on the
$\mu$-distortion); $r_L\approx 14\ \text{Gpc}$ is the distance to the
surface of last scattering; $3\cos(kr)e^{-k^2/k_D^2}$ comes from the
small-scale limit of the photon linear transfer function; and
$\langle\rangle_p$ denotes an average over the oscillation period. 

Correlating Eqs.~\eqref{eq:almT} and \eqref{eq:almM}, we find
\begin{align}\label{eq:CmuT-itgd}
  C^{\mu T}_l =& \frac{27}{20\pi^3} \int_0^\infty
 k_1^2dk_1\left[ e^{-2 k_1^2/k_D^2(z)}\right]_{z_f}^{z_i}\cr
&\times \int_0^\infty k^2 dk~ W\left(\frac{k}{k_s} \right)B_\zeta(k_1,
 k_2, k)
 j_l(k r_L) g_{Tl}(k),\cr
\end{align}
with $k_s\approx k_{D}(z_f)$.

In order to quantify how well we can measure $C^{\mu T}_l$ in real data,
we shall estimate the cumulative signal-to-noise ratio, $S/N$,
from 
\begin{align}\label{eq:Fischer-mat}
  \left(\frac{S}{N}\right)^2=\sum_l^{l_{\rm max}}(2l+1)\frac{{(C_l^{\mu T}})^2}{C_l^{TT}C_l^{\mu\mu,N}}.
\end{align}
Here, we have assumed that the temperature data on large scales are dominated by
the signal (which is already the case for the WMAP data), while the
$\mu$-type distortion data are dominated by noise.

\subsection{Estimating the noise level of $\mu$: absolutely calibrated
  experiments}
One can relate $\mu$ to a small change in the
CMB photon intensity, $I_\nu=(2h\nu^3/c)(e^{x+\mu}-1)^{-1}$, as
\begin{align}
 \delta I_\nu &= \left.\frac{\partial I_\nu}{\partial \mu}\right|_{\mu=0}\mu
= -\frac{2h\nu^3}{c^2}\frac{e^x}{(e^x-1)^2}\mu\cr
&= -2.70\times 10^{-18}~{\rm W~m^{-2}~Hz^{-1}~sr^{-1}}\frac{x^3 e^x}{(e^x-1)^2}\mu,\cr
\end{align}
where $x\equiv h\nu/(k_BT)=\nu/(56.80~{\rm GHz})$, for $T=2.725$~K. This
gives
\begin{equation}
 \frac{\mu}{10^{-8}}=-\frac{(e^x-1)^2}{x^3e^x}\frac{\delta
  I_\nu}{2.70\times 10^{-26}~{\rm W~m^{-2}~Hz^{-1}~sr^{-1}}},
\end{equation}
which can be used to estimate the noise level of $\mu$ from that of
$I_\nu$. A factor $\frac{(e^x-1)^2}{x^3e^x}$ is typically of order
unity: $\frac{(e^x-1)^2}{x^3e^x}=1.038$, 0.7307, 0.6568, and 0.8804 for
$\nu=60$, 100, 150, and 240~GHz, respectively.

For example, a proposed satellite experiment, PIXIE \cite{Kogut:2011xw},
is designed to have a typical noise level of $\delta I_\nu=4\times
10^{-24}~{\rm W~m^{-2}~Hz^{-1}~sr^{-1}}$ within each of 49152
equal-area pixels covering the full sky and each of 400 spectral
channels covering 30~GHz to 6~THz with a 15 GHz bandwidth. Averaging over
the full sky, PIXIE would reach $|\mu|\approx 0.5\times 10^{-8}$ at 100~GHz.

We shall assume that noise is white. The white noise level in the power
spectrum can be calculated as:
\begin{align}
 C_l^{\mu\mu,N}=&\left(\mbox{$1\sigma$-uncertainty in $\mu$ per
 pixel}\right)^2\cr 
&\times(\mbox{Solid angle of a pixel in units of
 steradians})\cr
&\times b_l^{-2},
\end{align}
where $b_l$ is the  the so-called beam transfer function, which is the
spherical harmonics coefficients of an experimental beam profile. For a
Gaussian beam with a full-width-at-half-maximum of $\theta_b$, $b_l$ is
given by
\begin{equation}
 b_l=\exp\left(-\frac{l^2\theta_b^2}{16\ln 2}\right).
\end{equation}
For PIXIE, we shall take the $1\sigma$-uncertainty in $\mu$
averaged over the full sky to be $10^{-8}$. Therefore, the white
noise level in the power spectrum is given by $4\pi\times 10^{-16}$
\cite{Pajer:2012vz}. Their beam has $\theta_b=1.6^\circ$, yielding
$C_l^{\mu\mu,N}=4\pi\times 10^{-16}\times e^{l^2/84^2}$.

\subsection{Estimating the noise level of $\mu$: relatively calibrated
  experiments}
\label{sec:estim-noise-level}
One must have an absolutely-calibrated experiment such as PIXIE in order
to measure a uniform $\mu$. However, an interesting implication of a
non-zero bispectrum in the squeezed configuration is that $\mu$ becomes
anisotropic. This induces a position-dependent temperature
fluctuation as $h\nu/(kT)\to h\nu/(kT)+\mu$, i.e.,
\begin{equation}
 T\to T(\hat{n})=\frac{2.725~{\rm K}}{1+\frac{\mu(\hat{n})}{x}}.
\end{equation}
The level of anisotropy is thus 
\begin{equation}
\frac{\delta T(\hat{n})}{T}\approx
-\frac{\delta\mu(\hat{n})}{x},
\end{equation}
where $\delta\mu$ is a fluctuating part of $\mu$, i.e.,
$\mu(\hat{n})=\bar{\mu}+\delta\mu(\hat{n})$.
Therefore, in principle, experiments which are calibrated to the CMB
dipole such as WMAP and Planck, as well as the proposed LiteBIRD
\footnote{{\sf http://cmbpol.kek.jp/litebird/}}, are also 
capable of measuring this effect by making a map of $\delta\mu(\hat{n})$
from the difference between temperature maps at two different
frequencies, $\nu_1$ and $\nu_2$. Then, the formula for the noise power
spectrum becomes
\begin{align}
 C_l^{\mu\mu,N}=&\left[\frac{\nu_1\nu_2/(\nu_1-\nu_2)}{56.80~{\rm GHz}}\right]^2\cr
&\times [(\mbox{$1\sigma$-uncertainty in $\delta T/T$ per
 pixel at $\nu_1$})^2\cr
&\phantom{\times}+(\mbox{$1\sigma$-uncertainty in $\delta T/T$ per
 pixel at $\nu_2$})^2]\cr 
&\times(\mbox{Solid angle of a pixel in units of
 steradians})\cr
&\times b_l^{-2}.
\label{eq:relativenoise}
\end{align}

According to the Planck Blue Book \cite{2006astro.ph..4069T}, the
expected sensitivities of Planck at 100 and 143~GHz are 
$\delta T/T=2.5\times
10^{-6}$ per $10'\times 10'$ pixel, and 
$\delta T/T=2.2\times
10^{-6}$ per $7.1'\times 7.1'$ pixel, respectively. The in-flight
performance then 
shows that the achieved noise level is 70\% of the expectation, and the
beam sizes at 100 and 143~GHz are 
$9.4'$ and $7.2'$, respectively \cite{PlanckHFICoreTeam:2011az}; thus, we
estimate Planck's sensitivity to $\delta\mu$ measured from maps at 100
and 143~GHz as $C_l^{\mu\mu,N}\approx 1.1\times 10^{-15}\times
e^{l^2/861^2}$. This is comparable to the above estimate for PIXIE,
which is based on the {\it absolute} measurement of the CMB
spectrum. 

However, as the sensitivity of PIXIE (and LiteBIRD) to CMB anisotropy is at
least an order of magnitude 
better than that of Planck, if we focus only on the
spatially-varying part of $\mu$ rather than the uniform part of $\mu$,
then it may be possible to increase the sensitivity to $C_l^{\mu T}$. The expected noise power spectrum is of order
 $C_l^{\mu\mu,N}\approx 10^{-17}$ or better, according to
 Eq.~(\ref{eq:relativenoise}). In other words, the 
 signal-to-noise of $C_l^{\mu T}$ can be improved by an order of
 magnitude as compared to the case where we look at absolute measurements of $\mu$.

In order to do this in practice, we must calibrate instruments at
different frequencies so that they have the equal response to the
thermal CMB. To estimate the required precision for calibration, let us
suppose that the 
response of one instrument at $\nu_1$ is different from that of another
at $\nu_2$ by $\epsilon$. Then, the difference between two maps at these
frequencies will yield $(\delta T/T)(\nu_1)-(\delta
T/T)(\nu_2)=\epsilon(\delta T/T)$, where $\delta T$ is the 
CMB anisotropy. This residual will be confused as a signal in
$\delta\mu$, such that
$\delta\mu=\left[\frac{\nu_1\nu_2/(\nu_1-\nu_2)}{56.80~{\rm GHz}}\right]\epsilon(\delta T/T)$ which, in
turn, will give a contamination of $C_l^{\mu T}$ given by 
\begin{eqnarray}
\nonumber
& &\frac{l(l+1)C_{l,\text{contamination}}^{\mu T}}{2\pi} \\
\nonumber
&=&2\times 10^{-10}\epsilon\left[\frac{\nu_1\nu_2/(\nu_1-\nu_2)}{56.80~{\rm GHz}}\right]\left[\frac{l(l+1)C_l^{TT}/2\pi}{2\times 10^{-10}}\right].\\
\end{eqnarray}
As we shall show below (also see \cite{Pajer:2012vz}), $C_l^{\mu T}$
from the local-form bispectrum is of order $l(l+1)|C_l^{\mu
T}|/(2\pi)\approx 4\times 10^{-17}f_{\rm NL}$. Therefore, the required precision
for the calibration is given by
\begin{equation}
\left[\frac{\nu_1\nu_2/(\nu_1-\nu_2)}{56.80~{\rm GHz}}\right]\epsilon\ll 2\times 10^{-7}f_{\rm NL}.
\end{equation}
For example, the calibration precision of the WMAP data is
$\epsilon=2\times 10^{-3}$ \cite{Hinshaw:2008kr,*Jarosik:2010iu}, and
thus the contamination of $C_l^{\mu T}$ due to the calibration mismatch
is negligible for $f_{\rm NL}\gg
10^4\left[\frac{\nu_1\nu_2/(\nu_1-\nu_2)}{56.80~{\rm GHz}}\right]$. To
search for a $C_l^{\mu T}$ signal, future CMB experiments such as
LiteBIRD may wish to place more emphasis on the relative calibration of
their instruments. Note that the signal from a modified initial state
will be much larger,  
$l(l+1)|C_l^{\mu T}|/(2\pi)\approx \frac{2\times
10^{-11}}{l}(\frac{\dot\phi^2}{2H^2}/0.01)$ (see Section \ref{sec:results-modif-init}),
so the required precision 
for the calibration can be relaxed greatly.

Another factor which may limit the utility of relatively calibrated
instruments is foreground contamination. In principle,
foreground contamination can be removed by using multi-frequency data, since the frequency spectrum of the foreground (roughly proportional to $\nu^{-3}$, $\nu^{-2}$, and $\nu^2$ for synchrotron, free-free, and dust emission, respectively) is
different from that of  $\mu$-type distortion ($\propto
\nu^{-1}$). Detailed study of foreground contamination is 
beyond the scope of this paper, but it is worth studying this in detail
once the required calibration precision is reached.

\subsection{Results for the local-form bispectrum}
Before we discuss $C_l^{\mu T}$ from the modified initial state effect,
let us discuss $C_l^{\mu T}$ from the local-form bispectrum. Using the
local-form bispectrum (Eq.~\eqref{eq:localform}) in the expression for
$C_l^{\mu T}$ (Eq.~\eqref{eq:CmuT-itgd}), we find 
\begin{align}
  C^{\mu T}_l =& \frac{81}{25\pi^3} f_{\rm NL}\int_0^\infty
 k_1^2dk_1\left[ e^{-2 k_1^2/k_D^2(z)}\right]_{z_f}^{z_i}P_\zeta(k_1)\cr
\times& \int_0^\infty k^2 dk~ W\left[\frac{k}{k_D(z_f)} \right]
P_\zeta(k)j_l(k r_L) g_{Tl}(k).
\end{align}
For a scale-invariant spectrum,
$P_\zeta=\frac{2\pi^2}{k^3}\Delta_\zeta^2$ with
$\Delta_\zeta^2=2.4\times 10^{-9}$, we find
\begin{align}
  C^{\mu T}_l =& \frac{324\pi}{25} f_{\rm
 NL}\Delta_\zeta^4\ln\left[\frac{k_D(z_i)}{k_D(z_f)}\right] \cr
&\times \int_0^\infty \frac{dk}{k}~ W\left[\frac{k}{k_D(z_f)} \right]j_l(k
 r_L) g_{Tl}(k). 
\end{align}
While we use the exact radiation transfer function calculated from the
linear Boltzmann code (except as indicated), it is instructive to obtain
an analytical 
expression for the Sachs-Wolfe limit, $g_{Tl}(k)\to -j_l(kr_L)/3$. As
the wavenumbers that are responsible for the Sachs-Wolfe regime are much
smaller than $k_D(z_f)$, one can approximate $W\left[\frac{k}{k_D(z_f)}
\right]\to 1$. We then find
\begin{eqnarray}
\nonumber
  C^{\mu T}_l &\to& -\frac{54\pi}{25} f_{\rm
 NL}\Delta_\zeta^4\ln\left[\frac{k_D(z_i)}{k_D(z_f)}\right]
 \frac{1}{l(l+1)}\\
&\approx& -3.5\times 10^{-17}f_{\rm NL}\times \frac{2\pi}{l(l+1)}.
\label{eq:localapp}
\end{eqnarray}
This result agrees with that obtained by \cite{Pajer:2012vz} up to a
factor of 3/4 recently found by \cite{Chluba:2012gq}. Therefore, on
large scales where the Sachs-Wolfe approximation is valid, the
cross-power spectrum is ``scale invariant,'' in a sense that
$l(l+1)C_l^{\mu T}=$constant. The overall sign is negative for a
positive $f_{\rm NL}$ because, for
a positive curvature perturbation $\zeta$, $\delta T/T=-\zeta/5$ is negative
whereas the fluctuation in $\delta\mu\propto f_{\rm NL}\zeta$ is
positive for a positive $f_{\rm NL}$.

\begin{figure}
  \centering
  \scalebox{0.48}{\includegraphics{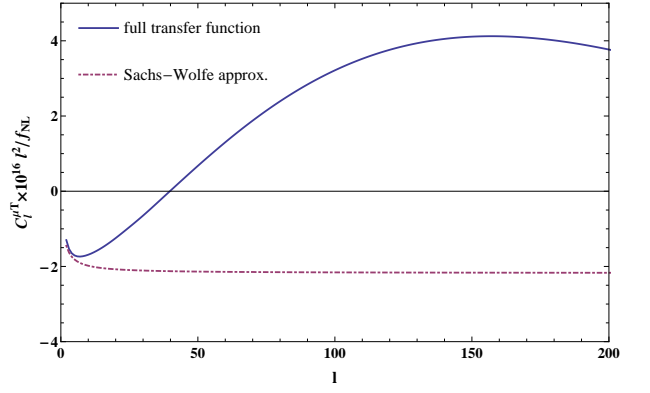}}  
  \caption{$C_l^{\mu T}$, the cross-power spectrum of the $\mu$-type distortion and the CMB temperature anisotropy, from the local-form
 bispectrum with $f_{\rm NL}=1$. The solid line shows $C_l^{\mu 
 T}$ using the full radiation transfer function, while the dot-dashed
 line shows it using the Sachs-Wolfe approximation. The amplitude of
 $C_l^{\mu T}$ is linearly proportional to $f_{\rm NL}$.}
  \label{fig:sw}
\end{figure}

How good is the Sachs-Wolfe approximation? In Figure~\ref{fig:sw}, we
compare the Sachs-Wolfe approximation with the exact calculation. We
find that the Sachs-Wolfe approximation breaks down at $l\approx 10$,
and the acoustic oscillation changes the sign of $C_l^{\mu T}$ at
$l\approx 40$. As $C_l^{\mu T}$ crosses zero at $l\approx 40$, we expect
the signal-to-noise ratio to grow more slowly with increasing multipole
than with the Sachs-Wolfe approximation.

\begin{figure}
  \centering
  \scalebox{0.48}{\includegraphics{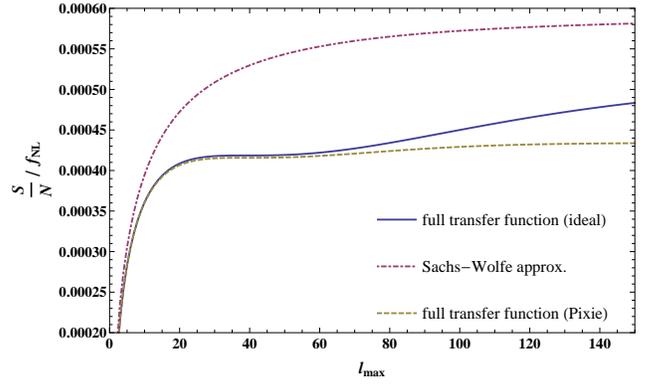}}
  \caption{Signal-to-noise ratio of $C_l^{\mu T}$ from the local-form
 bispectrum with $f_{\rm NL}=1$. The bottom two lines show $C_l^{\mu 
 T}$ using the full radiation transfer function, while the dot-dashed
 line shows it using the Sachs-Wolfe approximation. The solid, dashed line is for $\theta_b=0$ (ideal) and $1.6^\circ$ (PIXIE),
 respectively. The noise level is $C_l^{\mu\mu,N}=4\pi\times
 10^{-16}e^{l^2\theta_b^2/(8\ln 2)}$
 for all cases (i.e., the r.m.s. uncertainty of $\mu$ averaged over the
 full sky is 
 $10^{-8}$). The signal-to-noise is proportional to $f_{\rm NL}$ and is
 inversely proportional to the  r.m.s. uncertainty of $\mu$.}
  \label{fig:snlocal}
\end{figure}

In Figure~\ref{fig:snlocal}, we show the cumulative signal-to-noise
ratio of $C_l^{\mu T}$ from the local-form bispectrum as a function of
the maximum multipole, $l_{\rm max}$. We find that the Sachs-Wolfe
approximation overestimates the signal-to-noise ratio by about 40\%. 
Due to the sign change in $C_l^{\mu T}$, the signal-to-noise ratio does
not grow between $l_{\rm max}\approx 20$ and $80$.

For PIXIE's specification with $\theta_b=1.6^\circ$  and
$C_l^{\mu\mu,N}=4\pi\times 10^{-16}$, the signal-to-noise ratio reaches
$S/N= 4.3\times 10^{-4}f_{\rm NL}$. Therefore, PIXIE would be able to
see the signal if $f_{\rm NL}\gg 2300$. If PIXIE's detectors can be
calibrated so that the difference between maps at different frequencies
cancels the CMB anisotropy to the required precision, then $S/N$ can
improve by an order of magnitude. Reducing the beam size would not help
much because the signal-to-noise ratio  grows only logarithmically with
$l_{\rm max}$ \cite{Pajer:2012vz}.

\subsection{Results for the modified initial state bispectrum}
\label{sec:results-modif-init}
We now calculate $C_l^{\mu T}$ from the modified initial state
bispectrum. We start from
\begin{align}
  C^{\mu T}_l =& 
\frac{27}{20\pi^3} \frac{H^6}{\dot{\phi}^2}\xi(N)\int_0^\infty
 dk_1\left[ e^{-2 k_1^2/k_D^2(z)}\right]_{z_f}^{z_i}\cr
\times& \int_0^\infty \frac{dk}{k^2}~ W\left[\frac{k}{k_D(z_f)} \right]
j_l(k r_L) g_{Tl}(k)\kappa(k\eta_0)\cr
=& 
\frac{27}{40\sqrt{2}\pi^{5/2}}
 \frac{H^6}{\dot{\phi}^2}\xi(N)[k_D(z_i)-k_D(z_f)]\cr 
\times& \int_0^\infty \frac{dk}{k^2}~ W\left[\frac{k}{k_D(z_f)} \right]
j_l(k r_L) g_{Tl}(k)\kappa(k\eta_0),
\end{align}
where $\xi(N)$ is given by the second lines of
Eqs.~\eqref{eq:n_not_const_approx} and \eqref{eq:n_const_approx} and
$\kappa(k\eta_0)$ is given by the last lines of 
Eqs.~\eqref{eq:n_not_const_approx} and \eqref{eq:n_const_approx}. Once
again, a factor $H^6/\dot{\phi}^2$ will be calculated from the
normalization of $P_\zeta(k)$ for a given $\epsilon$, $N$, and $\theta$
(see Eq.~\eqref{eq:normalization}).

To obtain an order-of-magnitude estimation, let us set $\kappa=1$ and
take the Sachs-Wolfe limit. We then obtain
\begin{align}
  C^{\mu T}_l \to& 
-\frac{9}{40\sqrt{2}\pi^{5/2}}
 \frac{H^6}{\dot{\phi}^2}\xi(N)[k_D(z_i)-k_D(z_f)]\cr 
&\times \int_0^\infty \frac{dk}{k^2}~j_l^2(k r_L)\cr
=&
-\frac{9}{320\sqrt{2}\pi^{3/2}}
 \frac{H^6}{\dot{\phi}^2}\frac{\xi(N)[k_D(z_i)-k_D(z_f)]r_L}{\left(l+\frac32\right)\left(l+\frac12\right)\left(l-\frac12\right)}.\cr
\label{eq:nbdsw}
\end{align}
Therefore, $C^{\mu T}_l$ from the modified initial state falls as
$C^{\mu T}_l\propto l^{-3}$, which is faster than that from the local
form, $\propto l^{-2}$. However, the amplitude is proportional to
$k_D(z_i)r_L\approx 1.7\times 10^8$ instead of
$\ln[k_D(z_i)/k_D(z_f)]\approx 5.5$, which leads to a large 
amplification of the signal relative to the local form. 

\begin{figure}
  \centering
  \scalebox{0.48}{\includegraphics{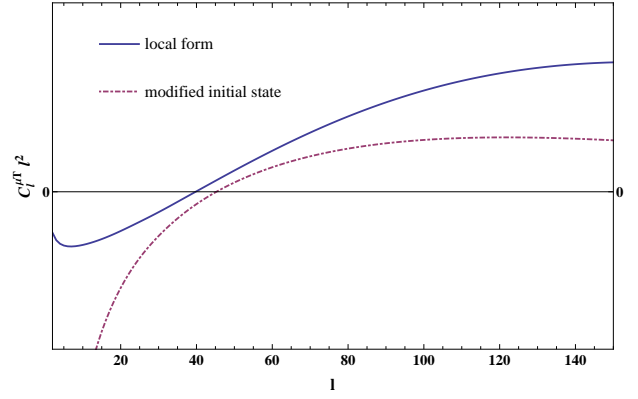}}  
  \caption{A comparison of the shapes of $C_l^{\mu T}$ (the cross-power spectra of the $\mu$-type distortion and the CMB temperature anisotropy) from the local-form bispectrum
 (solid line) and the modified state bispectrum (dot-dashed line). The amplitudes of the graphs are scaled so they can both appear in the same figure; therefore, the overall amplitude is arbitrary.}
  \label{fig:localvsnbd}
\end{figure}

In Figure~\ref{fig:localvsnbd}, we compare the shapes of $C_l^{\mu T}$ from the
local-form bispectrum (solid line) and from the modified initial state
bispectrum 
(dot-dashed line). As expected, for the low multipoles $l\lesssim 40$
(where $C_l^{\mu T}$ is 
negative), $C_l^{\mu T}$ from the modified initial
state is steeper (by a factor 
of $1/l$) than that from the local form, whereas for high
multipoles $l\gtrsim 40$ (where $C_l^{\mu T}$ is positive), $C_l^{\mu T}$ from the modified initial state is shallower (because it diminishes faster by $1/l$).

In order to better compare the modified initial state result
(Eq.~\eqref{eq:nbdsw}) with the local-form result
(Eq.~\eqref{eq:localapp}), let us use the power spectrum normalization,
$\Delta_\zeta^2=H^2(1+2N)/(8\pi^2\epsilon)$
(Eq.~\eqref{eq:normalization}), to rewrite Eq.~\eqref{eq:nbdsw} as
\begin{eqnarray}
\nonumber
  C^{\mu T}_l&=&
-\frac{9\pi^{5/2}}{10\sqrt{2}}\Delta_\zeta^4\epsilon\frac{\xi(N)}{(1+2N)^2}\frac{[k_D(z_i)-k_D(z_f)]r_L}{\left(l+\frac32\right)\left(l+\frac12\right)\left(l-\frac12\right)}\\
&\approx& \frac{-1.7\times
 10^{-11}}{l+\frac32}\frac{\epsilon}{10^{-2}}\frac{\xi(N)}{(1+2N)^2}\times
 \frac{2\pi}{l^2-\frac14}.
\label{eq:nbdapp}
\end{eqnarray}
We find the ratio
\begin{equation}
 \frac{C^{\mu T,\rm nBD}_l}{C_l^{\mu T,\rm local}}
\approx \frac{4.9\times 10^5}{(l+\frac32)f_{\rm NL}}
\frac{\epsilon}{10^{-2}}\frac{\xi(N)}{(1+2N)^2}\frac{l(l+1)}{l^2-\frac14}.
\label{eq:ratio}
\end{equation}
For single-field slow-roll inflation, $f_{\rm NL}\approx \epsilon\approx
10^{-2}$. Therefore, the ratio (Eq.~\eqref{eq:ratio}) is $\approx
5\times 10^7/l$.  

We can verify this ratio in a different way. Heuristically, for a modified initial state, the enhancement of the bispectrum in the squeezed configuration by a factor of $k_1/k\gg1$ becomes a enhancement of $C_l^{\mu T}$ by a factor of $k_D(z_i)r_L/l=k_D(z_i)/k_{\text{CMB},l}\approx 2\times 10^8/l$, i.e., the ratio of the acoustic damping wavenumber to the wavenumber generating the multipole $l$, giving close to the ratio above. Note that the biggest contribution to the signal will come from small multipoles (in particular the quadrupole $l=2$), so the signal boost relative to the local form is indeed large.

\begin{figure}
  \centering
  \scalebox{0.48}{\includegraphics{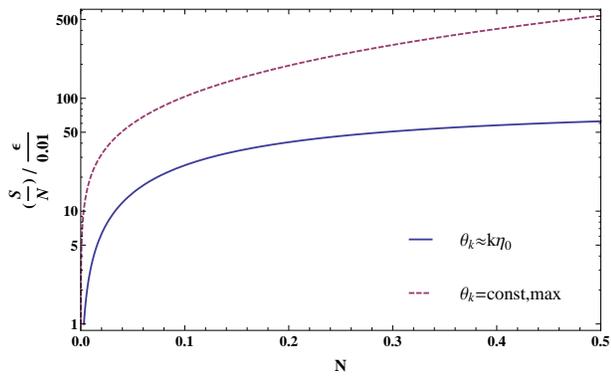}}  
  \caption{Signal-to-noise ratio of $C_l^{\mu T}$ from the modified
 initial state for $\epsilon=10^{-2}$, as a function of the occupation
 number $N$. 
The top and bottom
 lines are for $\theta_k\approx\text{const}$ (dashed) and $\theta_k\approx k
 \eta_0$ (solid), respectively.
The noise level is $C_l^{\mu\mu,N}=4\pi\times 10^{-16}e^{l^2/84^2}$
 for both  cases (i.e., the r.m.s. uncertainty of $\mu$ averaged over
 the full sky is 
 $10^{-8}$, and the beam size is $1.6^\circ$). The signal-to-noise ratio is
 proportional to $\epsilon$ and is 
 inversely proportional to the  r.m.s. uncertainty of $\mu$. }
  \label{fig:sig-to-noise_v_N}
\end{figure}

In Figure~\ref{fig:sig-to-noise_v_N}, we show the signal-to-noise ratio
of $C_l^{\mu T}$ for a modified initial state as a function of the
occupation number $N$. For PIXIE's specification with
$\theta_b=1.6^\circ$  and $C_l^{\mu\mu,N}=4\pi\times
10^{-16}e^{l^2\theta_b^2/(8\ln 2)}$, we find
a large signal-to-noise ratio: when the occupation number is $N\approx
0.5$, the signal-to-noise ratios are $S/N\approx 60(\epsilon/10^{-2})$ and
$500(\epsilon/10^{-2})$  for 
$\theta_k\approx k  \eta_0$ and $\theta_k\approx\text{const}$,
respectively. (Recall that, for the latter case, we have chosen a
constant $\theta_k$ so that it maximizes the signal.) This large
increase in the signal-to-noise ratio relative 
to the local-form bispectrum is consistent with the analytical estimate
given in Eq.~\eqref{eq:ratio}. 

Note that these calculations were done assuming $\eta_0$ was finite, i.e., we included the last lines of Eqs.~\eqref{eq:n_not_const_approx} and
 \eqref{eq:n_const_approx}. If we ignore the exponential terms in
 Eq.~\eqref{eq:nbd-bisp} by taking $\eta_0\to \eta_0+i\epsilon|\eta_0|$ for
 large $|\eta_0|$, then the signal-to-noise ratio goes down to about ten
 for $\theta_k\approx k  \eta_0$ and $N\approx 0.5$, which is still
 large enough for detection. Therefore, we 
 should be able to detect $C_l^{\mu T}$ in the PIXIE experiment unless
 $N$ were very small. 

 If the calibration of detectors at different
 frequencies meets the requirements (see
 Section~\ref{sec:estim-noise-level}), then PIXIE would be able to
 improve its signal-to-noise ratio for detecting $C_l^{\mu T}$ by an
 order of magnitude. Moreover, 
 if Planck's calibration  meets the requirement, Planck would be able
 to detect this signal. This merits further study.

However, one caveat should be mentioned, namely that these results
assume that the cut-off wavenumber, $k_{\text{cut}}$, lies above the
scales involved in the $\mu$-distortion. If $k_{\text{cut}}$ lay within
the $\mu$-distortion scales, this model could produce a measurable CMB
and LSS signal but have a smaller-than-expected $C_l^{\mu T}$.

\section{Conclusion}
\label{sec:discussion}

We have investigated phenomenological consequences of a modification of
 the  initial state of quantum fluctuations generated during
 single-field slow-roll inflation. In our model, the initial state is
 given by a Bogoliubov transformation on the standard Bunch-Davies
 initial vacuum state. A distinctive feature of this model
 is that the bispectrum of $\zeta$ in the squeezed configuration -- where
 one of the wavenumbers, $k$, is much smaller than the other two,
 i.e., $k\ll k_1\approx k_2$ -- is enhanced by a factor of $k_1/k$
 relative to the local-form bispectrum \cite{Agullo:2010ws}. This
 enhancement generates notable effects on the scale-dependent bias of LSS and on
 the $\mu$-type distortion of the black-body spectrum of CMB.

For LSS, the scale-dependent bias goes as $\bar{k}_1/k^3$ instead of
$1/k^2$, where $\bar{k}_1$ is a characteristic wavenumber corresponding
to  the short-wavelength mode in LSS for a given halo mass (Eq.~\eqref{eq:k1}).

For the $\mu$-type distortion, the squeezed
configuration bispectrum can make $\mu$ {\it anisotropic}, which can be
measured by cross-correlating a map of $\mu$ with a map of CMB temperature
anisotropy on large scales \cite{Pajer:2012vz}. The modified initial
state enhances power spectrum $C_l^{\mu T}$ of this cross-correlation by
a factor of $k_Dr_L$, which corrresponds to the ratio of the wavenumber of the
acoustic damping scale to the wavenumber measured by CMB temperature
anisotropy on large scales. We predict that an absolutely-calibrated
experiment such as PIXIE can detect $C_l^{\mu T}$ unless the occupation
number is much smaller than of order unity.

As this effect makes $\mu$ anisotropic, one may not even need an
absolutely-calibrated experiment. If detectors at different frequencies
are calibrated to have the same response to thermal CMB with the
sufficient precision, then relatively-calibrated experiments such as
Planck and LiteBIRD could detect this signal.

We acknowledge that our derivation of the bispectrum from a
modified initial state is limited by uncertainties about how to set
initial conditions and how to translate these conditions into a proper calculational framework. While we think that the calculations presented
in this paper capture plausible outcomes
of a modified initial state, more investigation on quantum field
theory with such a state is still necessary. That this model predicts
such interesting signatures in LSS and the CMB motivates further study.

Finally, while we have focused only on the bispectrum in the squeezed
configuration in this paper, this model also predicts a large bispectrum
in the folded limit, where the largest wavenumber is equal to the sum of
the other two wavenumbers, $k_1=k_2+k$
\cite{Chen:2006nt,Holman:2007na,Meerburg:2009ys}. The observational
signatures that we have discussed in this paper should come also with
the signal in the folded limit, which provides a powerful cross-check of
the nature of the detected signal. 

\acknowledgments
Note that a similar analysis on the effect of a modified initial state on the scale-dependent bias is presented in \cite{agullo-shandera}.

We would like to thank Xingang Chen, Daan Meerburg, and Enrico Pajer for helpful
discussions. This work is supported in part by NSF grant PHY-0758153.


\begin{thebibliography}{74}%
\makeatletter
\providecommand \@ifxundefined [1]{%
 \@ifx{#1\undefined}
}%
\providecommand \@ifnum [1]{%
 \ifnum #1\expandafter \@firstoftwo
 \else \expandafter \@secondoftwo
 \fi
}%
\providecommand \@ifx [1]{%
 \ifx #1\expandafter \@firstoftwo
 \else \expandafter \@secondoftwo
 \fi
}%
\providecommand \natexlab [1]{#1}%
\providecommand \enquote  [1]{``#1''}%
\providecommand \bibnamefont  [1]{#1}%
\providecommand \bibfnamefont [1]{#1}%
\providecommand \citenamefont [1]{#1}%
\providecommand \href@noop [0]{\@secondoftwo}%
\providecommand \href [0]{\begingroup \@sanitize@url \@href}%
\providecommand \@href[1]{\@@startlink{#1}\@@href}%
\providecommand \@@href[1]{\endgroup#1\@@endlink}%
\providecommand \@sanitize@url [0]{\catcode `\\12\catcode `\$12\catcode
  `\&12\catcode `\#12\catcode `\^12\catcode `\_12\catcode `\%12\relax}%
\providecommand \@@startlink[1]{}%
\providecommand \@@endlink[0]{}%
\providecommand \url  [0]{\begingroup\@sanitize@url \@url }%
\providecommand \@url [1]{\endgroup\@href {#1}{\urlprefix }}%
\providecommand \urlprefix  [0]{URL }%
\providecommand \Eprint [0]{\href }%
\providecommand \doibase [0]{http://dx.doi.org/}%
\providecommand \selectlanguage [0]{\@gobble}%
\providecommand \bibinfo  [0]{\@secondoftwo}%
\providecommand \bibfield  [0]{\@secondoftwo}%
\providecommand \translation [1]{[#1]}%
\providecommand \BibitemOpen [0]{}%
\providecommand \bibitemStop [0]{}%
\providecommand \bibitemNoStop [0]{.\EOS\space}%
\providecommand \EOS [0]{\spacefactor3000\relax}%
\providecommand \BibitemShut  [1]{\csname bibitem#1\endcsname}%
\let\auto@bib@innerbib\@empty
\bibitem [{\citenamefont {Bartolo}\ \emph {et~al.}(2004)\citenamefont
  {Bartolo}, \citenamefont {Komatsu}, \citenamefont {Matarrese},\ and\
  \citenamefont {Riotto}}]{bartolo/etal:2004}%
  \BibitemOpen
  \bibfield  {author} {\bibinfo {author} {\bibfnamefont {N.}~\bibnamefont
  {Bartolo}}, \bibinfo {author} {\bibfnamefont {E.}~\bibnamefont {Komatsu}},
  \bibinfo {author} {\bibfnamefont {S.}~\bibnamefont {Matarrese}}, \ and\
  \bibinfo {author} {\bibfnamefont {A.}~\bibnamefont {Riotto}},\ }\href@noop {}
  {\bibfield  {journal} {\bibinfo  {journal} {Phys. Rept.}\ }\textbf {\bibinfo
  {volume} {402}},\ \bibinfo {pages} {103} (\bibinfo {year} {2004})},\ \Eprint
  {http://arxiv.org/abs/astro-ph/0406398} {astro-ph/0406398} \BibitemShut
  {NoStop}%
\bibitem [{\citenamefont {{Komatsu}}\ \emph {et~al.}(2009)\citenamefont
  {{Komatsu}} \emph {et~al.}}]{2009astro2010S.158K}%
  \BibitemOpen
  \bibfield  {author} {\bibinfo {author} {\bibfnamefont {E.}~\bibnamefont
  {{Komatsu}}} \emph {et~al.},\ }in\ \href@noop {} {\emph {\bibinfo {booktitle}
  {astro2010: The Astronomy and Astrophysics Decadal Survey}}},\ \bibinfo
  {series} {Astronomy}, Vol.\ \bibinfo {volume} {2010}\ (\bibinfo {year}
  {2009})\ p.\ \bibinfo {pages} {158},\ \Eprint
  {http://arxiv.org/abs/0902.4759} {arXiv:0902.4759 [astro-ph.CO]} \BibitemShut
  {NoStop}%
\bibitem [{\citenamefont {Chen}(2010)}]{Chen:2010xka}%
  \BibitemOpen
  \bibfield  {author} {\bibinfo {author} {\bibfnamefont {X.}~\bibnamefont
  {Chen}},\ }\href {\doibase 10.1155/2010/638979} {\bibfield  {journal}
  {\bibinfo  {journal} {Adv.Astron.}\ }\textbf {\bibinfo {volume} {2010}},\
  \bibinfo {pages} {638979} (\bibinfo {year} {2010})},\ \Eprint
  {http://arxiv.org/abs/1002.1416} {arXiv:1002.1416 [astro-ph.CO]} \BibitemShut
  {NoStop}%
\bibitem [{\citenamefont {Weinberg}(2003)}]{Weinberg:2003}%
  \BibitemOpen
  \bibfield  {author} {\bibinfo {author} {\bibfnamefont {S.}~\bibnamefont
  {Weinberg}},\ }\href {\doibase 10.1103/PhysRevD.67.123504} {\bibfield
  {journal} {\bibinfo  {journal} {Phys.Rev.}\ }\textbf {\bibinfo {volume}
  {D67}},\ \bibinfo {pages} {123504} (\bibinfo {year} {2003})},\ \Eprint
  {http://arxiv.org/abs/astro-ph/0302326} {arXiv:astro-ph/0302326 [astro-ph]}
  \BibitemShut {NoStop}%
\bibitem [{\citenamefont {Komatsu}\ \emph {et~al.}(2011)\citenamefont {Komatsu}
  \emph {et~al.}}]{Komatsu:2010fb}%
  \BibitemOpen
  \bibfield  {author} {\bibinfo {author} {\bibfnamefont {E.}~\bibnamefont
  {Komatsu}} \emph {et~al.} (\bibinfo {collaboration} {WMAP Collaboration}),\
  }\href {\doibase 10.1088/0067-0049/192/2/18} {\bibfield  {journal} {\bibinfo
  {journal} {Astrophys.J.Suppl.}\ }\textbf {\bibinfo {volume} {192}},\ \bibinfo
  {pages} {18} (\bibinfo {year} {2011})},\ \Eprint
  {http://arxiv.org/abs/1001.4538} {arXiv:1001.4538 [astro-ph.CO]} \BibitemShut
  {NoStop}%
\bibitem [{\citenamefont {Dunkley}\ \emph {et~al.}(2011)\citenamefont {Dunkley}
  \emph {et~al.}}]{Dunkley:2010ge}%
  \BibitemOpen
  \bibfield  {author} {\bibinfo {author} {\bibfnamefont {J.}~\bibnamefont
  {Dunkley}} \emph {et~al.} (\bibinfo {collaboration} {ACT Collaboration}),\
  }\href {\doibase 10.1088/0004-637X/739/1/52} {\bibfield  {journal} {\bibinfo
  {journal} {Astrophys.J.}\ }\textbf {\bibinfo {volume} {739}},\ \bibinfo
  {pages} {52} (\bibinfo {year} {2011})},\ \Eprint
  {http://arxiv.org/abs/1009.0866} {arXiv:1009.0866 [astro-ph.CO]} \BibitemShut
  {NoStop}%
\bibitem [{\citenamefont {Keisler}\ \emph {et~al.}(2011)\citenamefont {Keisler}
  \emph {et~al.}}]{Keisler:2011aw}%
  \BibitemOpen
  \bibfield  {author} {\bibinfo {author} {\bibfnamefont {R.}~\bibnamefont
  {Keisler}} \emph {et~al.} (\bibinfo {collaboration} {SPT Collaboration}),\
  }\href {\doibase 10.1088/0004-637X/743/1/28} {\bibfield  {journal} {\bibinfo
  {journal} {Astrophys.J.}\ }\textbf {\bibinfo {volume} {743}},\ \bibinfo
  {pages} {28} (\bibinfo {year} {2011})},\ \Eprint
  {http://arxiv.org/abs/1105.3182} {arXiv:1105.3182 [astro-ph.CO]} \BibitemShut
  {NoStop}%
\bibitem [{\citenamefont {{Gangui}}\ \emph {et~al.}(1994)\citenamefont
  {{Gangui}}, \citenamefont {{Lucchin}}, \citenamefont {{Matarrese}},\ and\
  \citenamefont {{Mollerach}}}]{gangui/etal:1994}%
  \BibitemOpen
  \bibfield  {author} {\bibinfo {author} {\bibfnamefont {A.}~\bibnamefont
  {{Gangui}}}, \bibinfo {author} {\bibfnamefont {F.}~\bibnamefont {{Lucchin}}},
  \bibinfo {author} {\bibfnamefont {S.}~\bibnamefont {{Matarrese}}}, \ and\
  \bibinfo {author} {\bibfnamefont {S.}~\bibnamefont {{Mollerach}}},\
  }\href@noop {} {\bibfield  {journal} {\bibinfo  {journal} {Astrophys.J.}\
  }\textbf {\bibinfo {volume} {430}},\ \bibinfo {pages} {447} (\bibinfo {year}
  {1994})}\BibitemShut {NoStop}%
\bibitem [{\citenamefont {{Verde}}\ \emph {et~al.}(2000)\citenamefont
  {{Verde}}, \citenamefont {{Wang}}, \citenamefont {{Heavens}},\ and\
  \citenamefont {{Kamionkowski}}}]{verde/etal:2000}%
  \BibitemOpen
  \bibfield  {author} {\bibinfo {author} {\bibfnamefont {L.}~\bibnamefont
  {{Verde}}}, \bibinfo {author} {\bibfnamefont {L.}~\bibnamefont {{Wang}}},
  \bibinfo {author} {\bibfnamefont {A.~F.}\ \bibnamefont {{Heavens}}}, \ and\
  \bibinfo {author} {\bibfnamefont {M.}~\bibnamefont {{Kamionkowski}}},\
  }\href@noop {} {\bibfield  {journal} {\bibinfo  {journal}
  {Mon.Not.R.Astron.Soc.}\ }\textbf {\bibinfo {volume} {313}},\ \bibinfo
  {pages} {141} (\bibinfo {year} {2000})},\ \Eprint
  {http://arxiv.org/abs/arXiv:astro-ph/9906301} {arXiv:astro-ph/9906301}
  \BibitemShut {NoStop}%
\bibitem [{\citenamefont {Komatsu}\ and\ \citenamefont
  {Spergel}(2001)}]{Komatsu:2001rj}%
  \BibitemOpen
  \bibfield  {author} {\bibinfo {author} {\bibfnamefont {E.}~\bibnamefont
  {Komatsu}}\ and\ \bibinfo {author} {\bibfnamefont {D.~N.}\ \bibnamefont
  {Spergel}},\ }\href {\doibase 10.1103/PhysRevD.63.063002} {\bibfield
  {journal} {\bibinfo  {journal} {Phys. Rev.}\ }\textbf {\bibinfo {volume}
  {D63}},\ \bibinfo {pages} {063002} (\bibinfo {year} {2001})},\ \Eprint
  {http://arxiv.org/abs/astro-ph/0005036} {arXiv:astro-ph/0005036} \BibitemShut
  {NoStop}%
\bibitem [{\citenamefont {Maldacena}(2003)}]{Malda}%
  \BibitemOpen
  \bibfield  {author} {\bibinfo {author} {\bibfnamefont {J.~M.}\ \bibnamefont
  {Maldacena}},\ }\href@noop {} {\bibfield  {journal} {\bibinfo  {journal}
  {JHEP}\ }\textbf {\bibinfo {volume} {05}},\ \bibinfo {pages} {013} (\bibinfo
  {year} {2003})},\ \Eprint {http://arxiv.org/abs/astro-ph/0210603}
  {arXiv:astro-ph/0210603} \BibitemShut {NoStop}%
\bibitem [{\citenamefont {Acquaviva}\ \emph {et~al.}(2003)\citenamefont
  {Acquaviva}, \citenamefont {Bartolo}, \citenamefont {Matarrese},\ and\
  \citenamefont {Riotto}}]{Acquaviva:2002ud}%
  \BibitemOpen
  \bibfield  {author} {\bibinfo {author} {\bibfnamefont {V.}~\bibnamefont
  {Acquaviva}}, \bibinfo {author} {\bibfnamefont {N.}~\bibnamefont {Bartolo}},
  \bibinfo {author} {\bibfnamefont {S.}~\bibnamefont {Matarrese}}, \ and\
  \bibinfo {author} {\bibfnamefont {A.}~\bibnamefont {Riotto}},\ }\href
  {\doibase 10.1016/S0550-3213(03)00550-9} {\bibfield  {journal} {\bibinfo
  {journal} {Nucl.Phys.}\ }\textbf {\bibinfo {volume} {B667}},\ \bibinfo
  {pages} {119} (\bibinfo {year} {2003})},\ \Eprint
  {http://arxiv.org/abs/astro-ph/0209156} {arXiv:astro-ph/0209156 [astro-ph]}
  \BibitemShut {NoStop}%
\bibitem [{\citenamefont {Creminelli}\ and\ \citenamefont
  {Zaldarriaga}(2004)}]{creminelli04}%
  \BibitemOpen
  \bibfield  {author} {\bibinfo {author} {\bibfnamefont {P.}~\bibnamefont
  {Creminelli}}\ and\ \bibinfo {author} {\bibfnamefont {M.}~\bibnamefont
  {Zaldarriaga}},\ }\href {\doibase 10.1088/1475-7516/2004/10/006} {\bibfield
  {journal} {\bibinfo  {journal} {JCAP}\ }\textbf {\bibinfo {volume} {0410}},\
  \bibinfo {pages} {006} (\bibinfo {year} {2004})},\ \Eprint
  {http://arxiv.org/abs/astro-ph/0407059} {arXiv:astro-ph/0407059} \BibitemShut
  {NoStop}%
\bibitem [{\citenamefont {Creminelli}\ \emph
  {et~al.}(2011{\natexlab{a}})\citenamefont {Creminelli}, \citenamefont
  {D'Amico}, \citenamefont {Musso},\ and\ \citenamefont
  {Norena}}]{Creminelli:2011rh}%
  \BibitemOpen
  \bibfield  {author} {\bibinfo {author} {\bibfnamefont {P.}~\bibnamefont
  {Creminelli}}, \bibinfo {author} {\bibfnamefont {G.}~\bibnamefont {D'Amico}},
  \bibinfo {author} {\bibfnamefont {M.}~\bibnamefont {Musso}}, \ and\ \bibinfo
  {author} {\bibfnamefont {J.}~\bibnamefont {Norena}},\ }\href {\doibase
  10.1088/1475-7516/2011/11/038} {\bibfield  {journal} {\bibinfo  {journal}
  {JCAP}\ }\textbf {\bibinfo {volume} {1111}},\ \bibinfo {pages} {038}
  (\bibinfo {year} {2011}{\natexlab{a}})},\ \Eprint
  {http://arxiv.org/abs/1106.1462} {arXiv:1106.1462 [astro-ph.CO]} \BibitemShut
  {NoStop}%
\bibitem [{\citenamefont {Goldberg}\ and\ \citenamefont
  {Spergel}(1999)}]{Goldberg:1999xm}%
  \BibitemOpen
  \bibfield  {author} {\bibinfo {author} {\bibfnamefont {D.~M.}\ \bibnamefont
  {Goldberg}}\ and\ \bibinfo {author} {\bibfnamefont {D.~N.}\ \bibnamefont
  {Spergel}},\ }\href {\doibase 10.1103/PhysRevD.59.103002} {\bibfield
  {journal} {\bibinfo  {journal} {Phys.Rev.}\ }\textbf {\bibinfo {volume}
  {D59}},\ \bibinfo {pages} {103002} (\bibinfo {year} {1999})},\ \Eprint
  {http://arxiv.org/abs/astro-ph/9811251} {arXiv:astro-ph/9811251 [astro-ph]}
  \BibitemShut {NoStop}%
\bibitem [{\citenamefont {Verde}\ and\ \citenamefont
  {Spergel}(2002)}]{Verde:2002mu}%
  \BibitemOpen
  \bibfield  {author} {\bibinfo {author} {\bibfnamefont {L.}~\bibnamefont
  {Verde}}\ and\ \bibinfo {author} {\bibfnamefont {D.~N.}\ \bibnamefont
  {Spergel}},\ }\href {\doibase 10.1103/PhysRevD.65.043007} {\bibfield
  {journal} {\bibinfo  {journal} {Phys.Rev.}\ }\textbf {\bibinfo {volume}
  {D65}},\ \bibinfo {pages} {043007} (\bibinfo {year} {2002})},\ \Eprint
  {http://arxiv.org/abs/astro-ph/0108179} {arXiv:astro-ph/0108179 [astro-ph]}
  \BibitemShut {NoStop}%
\bibitem [{\citenamefont {Smith}\ and\ \citenamefont
  {Zaldarriaga}(2011)}]{Smith:2006ud}%
  \BibitemOpen
  \bibfield  {author} {\bibinfo {author} {\bibfnamefont {K.~M.}\ \bibnamefont
  {Smith}}\ and\ \bibinfo {author} {\bibfnamefont {M.}~\bibnamefont
  {Zaldarriaga}},\ }\href {\doibase 10.1111/j.1365-2966.2010.18175.x}
  {\bibfield  {journal} {\bibinfo  {journal} {Mon.Not.Roy.Astron.Soc.}\
  }\textbf {\bibinfo {volume} {417}},\ \bibinfo {pages} {2} (\bibinfo {year}
  {2011})},\ \Eprint {http://arxiv.org/abs/astro-ph/0612571}
  {arXiv:astro-ph/0612571 [astro-ph]} \BibitemShut {NoStop}%
\bibitem [{\citenamefont {Serra}\ and\ \citenamefont
  {Cooray}(2008)}]{Serra:2008wc}%
  \BibitemOpen
  \bibfield  {author} {\bibinfo {author} {\bibfnamefont {P.}~\bibnamefont
  {Serra}}\ and\ \bibinfo {author} {\bibfnamefont {A.}~\bibnamefont {Cooray}},\
  }\href {\doibase 10.1103/PhysRevD.77.107305} {\bibfield  {journal} {\bibinfo
  {journal} {Phys.Rev.}\ }\textbf {\bibinfo {volume} {D77}},\ \bibinfo {pages}
  {107305} (\bibinfo {year} {2008})},\ \Eprint {http://arxiv.org/abs/0801.3276}
  {arXiv:0801.3276 [astro-ph]} \BibitemShut {NoStop}%
\bibitem [{\citenamefont {Hanson}\ \emph {et~al.}(2009)\citenamefont {Hanson},
  \citenamefont {Smith}, \citenamefont {Challinor},\ and\ \citenamefont
  {Liguori}}]{Hanson:2009kg}%
  \BibitemOpen
  \bibfield  {author} {\bibinfo {author} {\bibfnamefont {D.}~\bibnamefont
  {Hanson}}, \bibinfo {author} {\bibfnamefont {K.~M.}\ \bibnamefont {Smith}},
  \bibinfo {author} {\bibfnamefont {A.}~\bibnamefont {Challinor}}, \ and\
  \bibinfo {author} {\bibfnamefont {M.}~\bibnamefont {Liguori}},\ }\href
  {\doibase 10.1103/PhysRevD.80.083004} {\bibfield  {journal} {\bibinfo
  {journal} {Phys.Rev.}\ }\textbf {\bibinfo {volume} {D80}},\ \bibinfo {pages}
  {083004} (\bibinfo {year} {2009})},\ \Eprint {http://arxiv.org/abs/0905.4732}
  {arXiv:0905.4732 [astro-ph.CO]} \BibitemShut {NoStop}%
\bibitem [{\citenamefont {Junk}\ and\ \citenamefont
  {Komatsu}(2012)}]{Junk:2012qt}%
  \BibitemOpen
  \bibfield  {author} {\bibinfo {author} {\bibfnamefont {V.}~\bibnamefont
  {Junk}}\ and\ \bibinfo {author} {\bibfnamefont {E.}~\bibnamefont {Komatsu}},\
  }\href@noop {} {\  (\bibinfo {year} {2012})},\ \Eprint
  {http://arxiv.org/abs/1204.3789} {arXiv:1204.3789 [astro-ph.CO]} \BibitemShut
  {NoStop}%
\bibitem [{\citenamefont {Nitta}\ \emph {et~al.}(2009)\citenamefont {Nitta},
  \citenamefont {Komatsu}, \citenamefont {Bartolo}, \citenamefont {Matarrese},\
  and\ \citenamefont {Riotto}}]{Nitta:2009jp}%
  \BibitemOpen
  \bibfield  {author} {\bibinfo {author} {\bibfnamefont {D.}~\bibnamefont
  {Nitta}}, \bibinfo {author} {\bibfnamefont {E.}~\bibnamefont {Komatsu}},
  \bibinfo {author} {\bibfnamefont {N.}~\bibnamefont {Bartolo}}, \bibinfo
  {author} {\bibfnamefont {S.}~\bibnamefont {Matarrese}}, \ and\ \bibinfo
  {author} {\bibfnamefont {A.}~\bibnamefont {Riotto}},\ }\href {\doibase
  10.1088/1475-7516/2009/05/014} {\bibfield  {journal} {\bibinfo  {journal}
  {JCAP}\ }\textbf {\bibinfo {volume} {0905}},\ \bibinfo {pages} {014}
  (\bibinfo {year} {2009})},\ \Eprint {http://arxiv.org/abs/0903.0894}
  {arXiv:0903.0894 [astro-ph.CO]} \BibitemShut {NoStop}%
\bibitem [{\citenamefont {Creminelli}\ \emph
  {et~al.}(2011{\natexlab{b}})\citenamefont {Creminelli}, \citenamefont
  {Pitrou},\ and\ \citenamefont {Vernizzi}}]{Creminelli:2011sq}%
  \BibitemOpen
  \bibfield  {author} {\bibinfo {author} {\bibfnamefont {P.}~\bibnamefont
  {Creminelli}}, \bibinfo {author} {\bibfnamefont {C.}~\bibnamefont {Pitrou}},
  \ and\ \bibinfo {author} {\bibfnamefont {F.}~\bibnamefont {Vernizzi}},\
  }\href {\doibase 10.1088/1475-7516/2011/11/025} {\bibfield  {journal}
  {\bibinfo  {journal} {JCAP}\ }\textbf {\bibinfo {volume} {1111}},\ \bibinfo
  {pages} {025} (\bibinfo {year} {2011}{\natexlab{b}})},\ \Eprint
  {http://arxiv.org/abs/1109.1822} {arXiv:1109.1822 [astro-ph.CO]} \BibitemShut
  {NoStop}%
\bibitem [{\citenamefont {Bartolo}\ \emph {et~al.}(2012)\citenamefont
  {Bartolo}, \citenamefont {Matarrese},\ and\ \citenamefont
  {Riotto}}]{Bartolo:2011wb}%
  \BibitemOpen
  \bibfield  {author} {\bibinfo {author} {\bibfnamefont {N.}~\bibnamefont
  {Bartolo}}, \bibinfo {author} {\bibfnamefont {S.}~\bibnamefont {Matarrese}},
  \ and\ \bibinfo {author} {\bibfnamefont {A.}~\bibnamefont {Riotto}},\ }\href
  {\doibase 10.1088/1475-7516/2012/02/017} {\bibfield  {journal} {\bibinfo
  {journal} {JCAP}\ }\textbf {\bibinfo {volume} {1202}},\ \bibinfo {pages}
  {017} (\bibinfo {year} {2012})},\ \Eprint {http://arxiv.org/abs/1109.2043}
  {arXiv:1109.2043 [astro-ph.CO]} \BibitemShut {NoStop}%
\bibitem [{\citenamefont {Agullo}\ and\ \citenamefont
  {Parker}(2011)}]{Agullo:2010ws}%
  \BibitemOpen
  \bibfield  {author} {\bibinfo {author} {\bibfnamefont {I.}~\bibnamefont
  {Agullo}}\ and\ \bibinfo {author} {\bibfnamefont {L.}~\bibnamefont
  {Parker}},\ }\href {\doibase 10.1103/PhysRevD.83.063526} {\bibfield
  {journal} {\bibinfo  {journal} {Phys.Rev.}\ }\textbf {\bibinfo {volume}
  {D83}},\ \bibinfo {pages} {063526} (\bibinfo {year} {2011})},\ \Eprint
  {http://arxiv.org/abs/1010.5766} {arXiv:1010.5766 [astro-ph.CO]} \BibitemShut
  {NoStop}%
\bibitem [{\citenamefont {Ganc}(2011)}]{Ganc:2011dy}%
  \BibitemOpen
  \bibfield  {author} {\bibinfo {author} {\bibfnamefont {J.}~\bibnamefont
  {Ganc}},\ }\href {\doibase 10.1103/PhysRevD.84.063514} {\bibfield  {journal}
  {\bibinfo  {journal} {Phys.Rev.}\ }\textbf {\bibinfo {volume} {D84}},\
  \bibinfo {pages} {063514} (\bibinfo {year} {2011})},\ \Eprint
  {http://arxiv.org/abs/1104.0244} {arXiv:1104.0244 [astro-ph.CO]} \BibitemShut
  {NoStop}%
\bibitem [{\citenamefont {Komatsu}\ \emph {et~al.}(2002)\citenamefont
  {Komatsu}, \citenamefont {Wandelt}, \citenamefont {Spergel}, \citenamefont
  {Banday},\ and\ \citenamefont {Gorski}}]{Komatsu:2001wu}%
  \BibitemOpen
  \bibfield  {author} {\bibinfo {author} {\bibfnamefont {E.}~\bibnamefont
  {Komatsu}}, \bibinfo {author} {\bibfnamefont {B.~D.}\ \bibnamefont
  {Wandelt}}, \bibinfo {author} {\bibfnamefont {D.~N.}\ \bibnamefont
  {Spergel}}, \bibinfo {author} {\bibfnamefont {A.~J.}\ \bibnamefont {Banday}},
  \ and\ \bibinfo {author} {\bibfnamefont {K.~M.}\ \bibnamefont {Gorski}},\
  }\href {\doibase 10.1086/337963} {\bibfield  {journal} {\bibinfo  {journal}
  {Astrophys.J.}\ }\textbf {\bibinfo {volume} {566}},\ \bibinfo {pages} {19}
  (\bibinfo {year} {2002})},\ \Eprint {http://arxiv.org/abs/astro-ph/0107605}
  {arXiv:astro-ph/0107605 [astro-ph]} \BibitemShut {NoStop}%
\bibitem [{\citenamefont {Komatsu}\ \emph {et~al.}(2003)\citenamefont {Komatsu}
  \emph {et~al.}}]{Komatsu:2003fd}%
  \BibitemOpen
  \bibfield  {author} {\bibinfo {author} {\bibfnamefont {E.}~\bibnamefont
  {Komatsu}} \emph {et~al.} (\bibinfo {collaboration} {WMAP Collaboration}),\
  }\href {\doibase 10.1086/377220} {\bibfield  {journal} {\bibinfo  {journal}
  {Astrophys.J.Suppl.}\ }\textbf {\bibinfo {volume} {148}},\ \bibinfo {pages}
  {119} (\bibinfo {year} {2003})},\ \Eprint
  {http://arxiv.org/abs/astro-ph/0302223} {arXiv:astro-ph/0302223 [astro-ph]}
  \BibitemShut {NoStop}%
\bibitem [{\citenamefont {Dalal}\ \emph {et~al.}(2008)\citenamefont {Dalal},
  \citenamefont {Dore}, \citenamefont {Huterer},\ and\ \citenamefont
  {Shirokov}}]{Dalal:2007cu}%
  \BibitemOpen
  \bibfield  {author} {\bibinfo {author} {\bibfnamefont {N.}~\bibnamefont
  {Dalal}}, \bibinfo {author} {\bibfnamefont {O.}~\bibnamefont {Dore}},
  \bibinfo {author} {\bibfnamefont {D.}~\bibnamefont {Huterer}}, \ and\
  \bibinfo {author} {\bibfnamefont {A.}~\bibnamefont {Shirokov}},\ }\href
  {\doibase 10.1103/PhysRevD.77.123514} {\bibfield  {journal} {\bibinfo
  {journal} {Phys.Rev.}\ }\textbf {\bibinfo {volume} {D77}},\ \bibinfo {pages}
  {123514} (\bibinfo {year} {2008})},\ \Eprint {http://arxiv.org/abs/0710.4560}
  {arXiv:0710.4560 [astro-ph]} \BibitemShut {NoStop}%
\bibitem [{\citenamefont {Slosar}\ \emph {et~al.}(2008)\citenamefont {Slosar},
  \citenamefont {Hirata}, \citenamefont {Seljak}, \citenamefont {Ho},\ and\
  \citenamefont {Padmanabhan}}]{Slosar:2008hx}%
  \BibitemOpen
  \bibfield  {author} {\bibinfo {author} {\bibfnamefont {A.}~\bibnamefont
  {Slosar}}, \bibinfo {author} {\bibfnamefont {C.}~\bibnamefont {Hirata}},
  \bibinfo {author} {\bibfnamefont {U.}~\bibnamefont {Seljak}}, \bibinfo
  {author} {\bibfnamefont {S.}~\bibnamefont {Ho}}, \ and\ \bibinfo {author}
  {\bibfnamefont {N.}~\bibnamefont {Padmanabhan}},\ }\href {\doibase
  10.1088/1475-7516/2008/08/031} {\bibfield  {journal} {\bibinfo  {journal}
  {JCAP}\ }\textbf {\bibinfo {volume} {0808}},\ \bibinfo {pages} {031}
  (\bibinfo {year} {2008})},\ \Eprint {http://arxiv.org/abs/0805.3580}
  {arXiv:0805.3580 [astro-ph]} \BibitemShut {NoStop}%
\bibitem [{\citenamefont {Matarrese}\ and\ \citenamefont
  {Verde}(2008)}]{Matarrese:2008nc}%
  \BibitemOpen
  \bibfield  {author} {\bibinfo {author} {\bibfnamefont {S.}~\bibnamefont
  {Matarrese}}\ and\ \bibinfo {author} {\bibfnamefont {L.}~\bibnamefont
  {Verde}},\ }\href {\doibase 10.1086/587840} {\bibfield  {journal} {\bibinfo
  {journal} {Astrophys.J.}\ }\textbf {\bibinfo {volume} {677}},\ \bibinfo
  {pages} {L77} (\bibinfo {year} {2008})},\ \Eprint
  {http://arxiv.org/abs/0801.4826} {arXiv:0801.4826 [astro-ph]} \BibitemShut
  {NoStop}%
\bibitem [{\citenamefont {Pajer}\ and\ \citenamefont
  {Zaldarriaga}(2012)}]{Pajer:2012vz}%
  \BibitemOpen
  \bibfield  {author} {\bibinfo {author} {\bibfnamefont {E.}~\bibnamefont
  {Pajer}}\ and\ \bibinfo {author} {\bibfnamefont {M.}~\bibnamefont
  {Zaldarriaga}},\ }\href@noop {} {\  (\bibinfo {year} {2012})},\ \Eprint
  {http://arxiv.org/abs/1201.5375} {arXiv:1201.5375 [astro-ph.CO]} \BibitemShut
  {NoStop}%
\bibitem [{\citenamefont {Komatsu}\ \emph {et~al.}(2009)\citenamefont {Komatsu}
  \emph {et~al.}}]{Komatsu:2008hk}%
  \BibitemOpen
  \bibfield  {author} {\bibinfo {author} {\bibfnamefont {E.}~\bibnamefont
  {Komatsu}} \emph {et~al.} (\bibinfo {collaboration} {WMAP Collaboration}),\
  }\href {\doibase 10.1088/0067-0049/180/2/330} {\bibfield  {journal} {\bibinfo
   {journal} {Astrophys.J.Suppl.}\ }\textbf {\bibinfo {volume} {180}},\
  \bibinfo {pages} {330} (\bibinfo {year} {2009})},\ \Eprint
  {http://arxiv.org/abs/0803.0547} {arXiv:0803.0547 [astro-ph]} \BibitemShut
  {NoStop}%
\bibitem [{\citenamefont {Kundu}(2012)}]{Kundu:Iftn-General-Init}%
  \BibitemOpen
  \bibfield  {author} {\bibinfo {author} {\bibfnamefont {S.}~\bibnamefont
  {Kundu}},\ }\href {\doibase 10.1088/1475-7516/2012/02/005} {\bibfield
  {journal} {\bibinfo  {journal} {JCAP}\ }\textbf {\bibinfo {volume} {1202}},\
  \bibinfo {pages} {005} (\bibinfo {year} {2012})},\ \Eprint
  {http://arxiv.org/abs/1110.4688} {arXiv:1110.4688 [astro-ph.CO]} \BibitemShut
  {NoStop}%
\bibitem [{\citenamefont {{Anderson}}\ \emph {et~al.}(2005)\citenamefont
  {{Anderson}}, \citenamefont {{Molina-Par{\'{\i}}s}},\ and\ \citenamefont
  {{Mottola}}}]{Anderson-Molina-Mottola}%
  \BibitemOpen
  \bibfield  {author} {\bibinfo {author} {\bibfnamefont {P.~R.}\ \bibnamefont
  {{Anderson}}}, \bibinfo {author} {\bibfnamefont {C.}~\bibnamefont
  {{Molina-Par{\'{\i}}s}}}, \ and\ \bibinfo {author} {\bibfnamefont
  {E.}~\bibnamefont {{Mottola}}},\ }\href {\doibase 10.1103/PhysRevD.72.043515}
  {\bibfield  {journal} {\bibinfo  {journal} {Phys.Rev.}\ }\textbf {\bibinfo
  {volume} {72}},\ \bibinfo {eid} {043515} (\bibinfo {year} {2005})},\ \Eprint
  {http://arxiv.org/abs/arXiv:hep-th/0504134} {arXiv:hep-th/0504134}
  \BibitemShut {NoStop}%
\bibitem [{\citenamefont {{Boyanovsky}}\ \emph {et~al.}(2006)\citenamefont
  {{Boyanovsky}}, \citenamefont {{de Vega}},\ and\ \citenamefont
  {{Sanchez}}}]{Boyanovsky:2006qi}%
  \BibitemOpen
  \bibfield  {author} {\bibinfo {author} {\bibfnamefont {D.}~\bibnamefont
  {{Boyanovsky}}}, \bibinfo {author} {\bibfnamefont {H.~J.}\ \bibnamefont {{de
  Vega}}}, \ and\ \bibinfo {author} {\bibfnamefont {N.~G.}\ \bibnamefont
  {{Sanchez}}},\ }\href {\doibase 10.1103/PhysRevD.74.123006} {\bibfield
  {journal} {\bibinfo  {journal} {\prd}\ }\textbf {\bibinfo {volume} {74}},\
  \bibinfo {eid} {123006} (\bibinfo {year} {2006})},\ \Eprint
  {http://arxiv.org/abs/arXiv:astro-ph/0607508} {arXiv:astro-ph/0607508}
  \BibitemShut {NoStop}%
\bibitem [{\citenamefont {Holman}\ and\ \citenamefont
  {Tolley}(2008)}]{Holman:2007na}%
  \BibitemOpen
  \bibfield  {author} {\bibinfo {author} {\bibfnamefont {R.}~\bibnamefont
  {Holman}}\ and\ \bibinfo {author} {\bibfnamefont {A.~J.}\ \bibnamefont
  {Tolley}},\ }\href {\doibase 10.1088/1475-7516/2008/05/001} {\bibfield
  {journal} {\bibinfo  {journal} {JCAP}\ }\textbf {\bibinfo {volume} {0805}},\
  \bibinfo {pages} {001} (\bibinfo {year} {2008})},\ \Eprint
  {http://arxiv.org/abs/0710.1302} {arXiv:0710.1302 [hep-th]} \BibitemShut
  {NoStop}%
\bibitem [{\citenamefont {Meerburg}\ \emph {et~al.}(2009)\citenamefont
  {Meerburg}, \citenamefont {van~der Schaar},\ and\ \citenamefont
  {Corasaniti}}]{Meerburg:2009ys}%
  \BibitemOpen
  \bibfield  {author} {\bibinfo {author} {\bibfnamefont {P.~D.}\ \bibnamefont
  {Meerburg}}, \bibinfo {author} {\bibfnamefont {J.~P.}\ \bibnamefont {van~der
  Schaar}}, \ and\ \bibinfo {author} {\bibfnamefont {P.~S.}\ \bibnamefont
  {Corasaniti}},\ }\href {\doibase 10.1088/1475-7516/2009/05/018} {\bibfield
  {journal} {\bibinfo  {journal} {JCAP}\ }\textbf {\bibinfo {volume} {0905}},\
  \bibinfo {pages} {018} (\bibinfo {year} {2009})},\ \Eprint
  {http://arxiv.org/abs/0901.4044} {arXiv:0901.4044 [hep-th]} \BibitemShut
  {NoStop}%
\bibitem [{\citenamefont {Meerburg}\ \emph {et~al.}(2010)\citenamefont
  {Meerburg}, \citenamefont {van~der Schaar},\ and\ \citenamefont
  {Jackson}}]{Meerburg:2009fi}%
  \BibitemOpen
  \bibfield  {author} {\bibinfo {author} {\bibfnamefont {P.}~\bibnamefont
  {Meerburg}}, \bibinfo {author} {\bibfnamefont {J.~P.}\ \bibnamefont {van~der
  Schaar}}, \ and\ \bibinfo {author} {\bibfnamefont {M.~G.}\ \bibnamefont
  {Jackson}},\ }\href {\doibase 10.1088/1475-7516/2010/02/001} {\bibfield
  {journal} {\bibinfo  {journal} {JCAP}\ }\textbf {\bibinfo {volume} {1002}},\
  \bibinfo {pages} {001} (\bibinfo {year} {2010})},\ \Eprint
  {http://arxiv.org/abs/0910.4986} {arXiv:0910.4986 [hep-th]} \BibitemShut
  {NoStop}%
\bibitem [{\citenamefont {Ashoorioon}\ and\ \citenamefont
  {Shiu}(2011)}]{Ashoorioon:2010xg}%
  \BibitemOpen
  \bibfield  {author} {\bibinfo {author} {\bibfnamefont {A.}~\bibnamefont
  {Ashoorioon}}\ and\ \bibinfo {author} {\bibfnamefont {G.}~\bibnamefont
  {Shiu}},\ }\href {\doibase 10.1088/1475-7516/2011/03/025} {\bibfield
  {journal} {\bibinfo  {journal} {JCAP}\ }\textbf {\bibinfo {volume} {1103}},\
  \bibinfo {pages} {025} (\bibinfo {year} {2011})},\ \Eprint
  {http://arxiv.org/abs/1012.3392} {arXiv:1012.3392 [astro-ph.CO]} \BibitemShut
  {NoStop}%
\bibitem [{\citenamefont {Schalm}\ \emph {et~al.}(2004)\citenamefont {Schalm},
  \citenamefont {Shiu},\ and\ \citenamefont {van~der Schaar}}]{Schalm:2004qk}%
  \BibitemOpen
  \bibfield  {author} {\bibinfo {author} {\bibfnamefont {K.}~\bibnamefont
  {Schalm}}, \bibinfo {author} {\bibfnamefont {G.}~\bibnamefont {Shiu}}, \ and\
  \bibinfo {author} {\bibfnamefont {J.~P.}\ \bibnamefont {van~der Schaar}},\
  }\href {\doibase 10.1088/1126-6708/2004/04/076} {\bibfield  {journal}
  {\bibinfo  {journal} {JHEP}\ }\textbf {\bibinfo {volume} {0404}},\ \bibinfo
  {pages} {076} (\bibinfo {year} {2004})},\ \Eprint
  {http://arxiv.org/abs/hep-th/0401164} {arXiv:hep-th/0401164 [hep-th]}
  \BibitemShut {NoStop}%
\bibitem [{\citenamefont {{Greene}}\ \emph {et~al.}(2005)\citenamefont
  {{Greene}}, \citenamefont {{Schalm}}, \citenamefont {{van der Schaar}},\ and\
  \citenamefont {{Shiu}}}]{Greene-BEFT}%
  \BibitemOpen
  \bibfield  {author} {\bibinfo {author} {\bibfnamefont {B.}~\bibnamefont
  {{Greene}}}, \bibinfo {author} {\bibfnamefont {K.}~\bibnamefont {{Schalm}}},
  \bibinfo {author} {\bibfnamefont {J.~P.}\ \bibnamefont {{van der Schaar}}}, \
  and\ \bibinfo {author} {\bibfnamefont {G.}~\bibnamefont {{Shiu}}},\ }in\
  \href@noop {} {\emph {\bibinfo {booktitle} {22nd Texas Symposium on
  Relativistic Astrophysics}}},\ \bibinfo {editor} {edited by\ \bibinfo
  {editor} {\bibnamefont {{P.~Chen, E.~Bloom, G.~Madejski, \& V.~Patrosian}}}}\
  (\bibinfo {year} {2005})\ pp.\ \bibinfo {pages} {1--8},\ \Eprint
  {http://arxiv.org/abs/arXiv:astro-ph/0503458} {arXiv:astro-ph/0503458}
  \BibitemShut {NoStop}%
\bibitem [{\citenamefont {{Easther}}\ \emph {et~al.}(2005)\citenamefont
  {{Easther}}, \citenamefont {{Kinney}},\ and\ \citenamefont
  {{Peiris}}}]{Easther:BEFT-v-NPH}%
  \BibitemOpen
  \bibfield  {author} {\bibinfo {author} {\bibfnamefont {R.}~\bibnamefont
  {{Easther}}}, \bibinfo {author} {\bibfnamefont {W.~H.}\ \bibnamefont
  {{Kinney}}}, \ and\ \bibinfo {author} {\bibfnamefont {H.}~\bibnamefont
  {{Peiris}}},\ }\href {\doibase 10.1088/1475-7516/2005/08/001} {\bibfield
  {journal} {\bibinfo  {journal} {JCAP}\ }\textbf {\bibinfo {volume} {8}},\
  \bibinfo {pages} {1} (\bibinfo {year} {2005})},\ \Eprint
  {http://arxiv.org/abs/arXiv:astro-ph/0505426} {arXiv:astro-ph/0505426}
  \BibitemShut {NoStop}%
\bibitem [{\citenamefont {Chen}\ \emph {et~al.}(2007)\citenamefont {Chen},
  \citenamefont {Huang}, \citenamefont {Kachru},\ and\ \citenamefont
  {Shiu}}]{Chen:2006nt}%
  \BibitemOpen
  \bibfield  {author} {\bibinfo {author} {\bibfnamefont {X.}~\bibnamefont
  {Chen}}, \bibinfo {author} {\bibfnamefont {M.-x.}\ \bibnamefont {Huang}},
  \bibinfo {author} {\bibfnamefont {S.}~\bibnamefont {Kachru}}, \ and\ \bibinfo
  {author} {\bibfnamefont {G.}~\bibnamefont {Shiu}},\ }\href {\doibase
  10.1088/1475-7516/2007/01/002} {\bibfield  {journal} {\bibinfo  {journal}
  {JCAP}\ }\textbf {\bibinfo {volume} {0701}},\ \bibinfo {pages} {002}
  (\bibinfo {year} {2007})},\ \Eprint {http://arxiv.org/abs/hep-th/0605045}
  {arXiv:hep-th/0605045 [hep-th]} \BibitemShut {NoStop}%
\bibitem [{\citenamefont {Sefusatti}\ and\ \citenamefont
  {Komatsu}(2007)}]{Sefusatti:2007ih}%
  \BibitemOpen
  \bibfield  {author} {\bibinfo {author} {\bibfnamefont {E.}~\bibnamefont
  {Sefusatti}}\ and\ \bibinfo {author} {\bibfnamefont {E.}~\bibnamefont
  {Komatsu}},\ }\href {\doibase 10.1103/PhysRevD.76.083004} {\bibfield
  {journal} {\bibinfo  {journal} {Phys.Rev.}\ }\textbf {\bibinfo {volume}
  {D76}},\ \bibinfo {pages} {083004} (\bibinfo {year} {2007})},\ \Eprint
  {http://arxiv.org/abs/0705.0343} {arXiv:0705.0343 [astro-ph]} \BibitemShut
  {NoStop}%
\bibitem [{\citenamefont {Grinstein}\ and\ \citenamefont
  {Wise}(1986)}]{Grinstein:1986en}%
  \BibitemOpen
  \bibfield  {author} {\bibinfo {author} {\bibfnamefont {B.}~\bibnamefont
  {Grinstein}}\ and\ \bibinfo {author} {\bibfnamefont {M.~B.}\ \bibnamefont
  {Wise}},\ }\href {\doibase 10.1086/164660} {\bibfield  {journal} {\bibinfo
  {journal} {Astrophys.J.}\ }\textbf {\bibinfo {volume} {310}},\ \bibinfo
  {pages} {19} (\bibinfo {year} {1986})}\BibitemShut {NoStop}%
\bibitem [{\citenamefont {Matarrese}\ \emph {et~al.}(1986)\citenamefont
  {Matarrese}, \citenamefont {Lucchin},\ and\ \citenamefont
  {Bonometto}}]{Matarrese:1986et}%
  \BibitemOpen
  \bibfield  {author} {\bibinfo {author} {\bibfnamefont {S.}~\bibnamefont
  {Matarrese}}, \bibinfo {author} {\bibfnamefont {F.}~\bibnamefont {Lucchin}},
  \ and\ \bibinfo {author} {\bibfnamefont {S.~A.}\ \bibnamefont {Bonometto}},\
  }\href@noop {} {\bibfield  {journal} {\bibinfo  {journal} {Astrophys.J.}\
  }\textbf {\bibinfo {volume} {310}},\ \bibinfo {pages} {L21} (\bibinfo {year}
  {1986})}\BibitemShut {NoStop}%
\bibitem [{\citenamefont {Kaiser}(1984)}]{Kaiser:1984sw}%
  \BibitemOpen
  \bibfield  {author} {\bibinfo {author} {\bibfnamefont {N.}~\bibnamefont
  {Kaiser}},\ }\href@noop {} {\bibfield  {journal} {\bibinfo  {journal}
  {Astrophys.J.}\ }\textbf {\bibinfo {volume} {284}},\ \bibinfo {pages} {L9}
  (\bibinfo {year} {1984})}\BibitemShut {NoStop}%
\bibitem [{\citenamefont {Desjacques}\ \emph
  {et~al.}(2011{\natexlab{a}})\citenamefont {Desjacques}, \citenamefont
  {Jeong},\ and\ \citenamefont {Schmidt}}]{Desjacques:2011mq}%
  \BibitemOpen
  \bibfield  {author} {\bibinfo {author} {\bibfnamefont {V.}~\bibnamefont
  {Desjacques}}, \bibinfo {author} {\bibfnamefont {D.}~\bibnamefont {Jeong}}, \
  and\ \bibinfo {author} {\bibfnamefont {F.}~\bibnamefont {Schmidt}},\ }\href
  {\doibase 10.1103/PhysRevD.84.063512} {\bibfield  {journal} {\bibinfo
  {journal} {Phys.Rev.}\ }\textbf {\bibinfo {volume} {D84}},\ \bibinfo {pages}
  {063512} (\bibinfo {year} {2011}{\natexlab{a}})},\ \Eprint
  {http://arxiv.org/abs/1105.3628} {arXiv:1105.3628 [astro-ph.CO]} \BibitemShut
  {NoStop}%
\bibitem [{\citenamefont {Desjacques}\ \emph
  {et~al.}(2011{\natexlab{b}})\citenamefont {Desjacques}, \citenamefont
  {Jeong},\ and\ \citenamefont {Schmidt}}]{Desjacques:2011jb}%
  \BibitemOpen
  \bibfield  {author} {\bibinfo {author} {\bibfnamefont {V.}~\bibnamefont
  {Desjacques}}, \bibinfo {author} {\bibfnamefont {D.}~\bibnamefont {Jeong}}, \
  and\ \bibinfo {author} {\bibfnamefont {F.}~\bibnamefont {Schmidt}},\ }\href
  {\doibase 10.1103/PhysRevD.84.061301} {\bibfield  {journal} {\bibinfo
  {journal} {Phys.Rev.}\ }\textbf {\bibinfo {volume} {D84}},\ \bibinfo {pages}
  {061301} (\bibinfo {year} {2011}{\natexlab{b}})},\ \Eprint
  {http://arxiv.org/abs/1105.3476} {arXiv:1105.3476 [astro-ph.CO]} \BibitemShut
  {NoStop}%
\bibitem [{\citenamefont {Chialva}(2011)}]{Chialva:2011hc}%
  \BibitemOpen
  \bibfield  {author} {\bibinfo {author} {\bibfnamefont {D.}~\bibnamefont
  {Chialva}},\ }\href@noop {} {\bibfield  {journal} {\bibinfo  {journal} {ArXiv
  e-prints}\ } (\bibinfo {year} {2011})},\ \Eprint
  {http://arxiv.org/abs/1108.4203} {arXiv:1108.4203 [astro-ph.CO]} \BibitemShut
  {NoStop}%
\bibitem [{\citenamefont {Sunyaev}\ and\ \citenamefont
  {Zeldovich}(1970)}]{Sunyaev:1970er}%
  \BibitemOpen
  \bibfield  {author} {\bibinfo {author} {\bibfnamefont {R.}~\bibnamefont
  {Sunyaev}}\ and\ \bibinfo {author} {\bibfnamefont {Y.}~\bibnamefont
  {Zeldovich}},\ }\href@noop {} {\bibfield  {journal} {\bibinfo  {journal}
  {Astrophys.Space Sci.}\ }\textbf {\bibinfo {volume} {7}},\ \bibinfo {pages}
  {20} (\bibinfo {year} {1970})}\BibitemShut {NoStop}%
\bibitem [{\citenamefont {{Sunyaev}}\ and\ \citenamefont
  {{Zeldovich}}(1970)}]{1970ApSS...9..368S}%
  \BibitemOpen
  \bibfield  {author} {\bibinfo {author} {\bibfnamefont {R.~A.}\ \bibnamefont
  {{Sunyaev}}}\ and\ \bibinfo {author} {\bibfnamefont {Y.~B.}\ \bibnamefont
  {{Zeldovich}}},\ }\href {\doibase 10.1007/BF00649577} {\bibfield  {journal}
  {\bibinfo  {journal} {Astrophys.Space Sci.}\ }\textbf {\bibinfo {volume}
  {9}},\ \bibinfo {pages} {368} (\bibinfo {year} {1970})}\BibitemShut {NoStop}%
\bibitem [{\citenamefont {{Daly}}(1991)}]{1991ApJ...371...14D}%
  \BibitemOpen
  \bibfield  {author} {\bibinfo {author} {\bibfnamefont {R.~A.}\ \bibnamefont
  {{Daly}}},\ }\href {\doibase 10.1086/169866} {\bibfield  {journal} {\bibinfo
  {journal} {Astrophys.J.}\ }\textbf {\bibinfo {volume} {371}},\ \bibinfo
  {pages} {14} (\bibinfo {year} {1991})}\BibitemShut {NoStop}%
\bibitem [{\citenamefont {Hu}\ and\ \citenamefont {Silk}(1993)}]{Hu:1992dc}%
  \BibitemOpen
  \bibfield  {author} {\bibinfo {author} {\bibfnamefont {W.}~\bibnamefont
  {Hu}}\ and\ \bibinfo {author} {\bibfnamefont {J.}~\bibnamefont {Silk}},\
  }\href {\doibase 10.1103/PhysRevD.48.485} {\bibfield  {journal} {\bibinfo
  {journal} {Phys.Rev.}\ }\textbf {\bibinfo {volume} {D48}},\ \bibinfo {pages}
  {485} (\bibinfo {year} {1993})}\BibitemShut {NoStop}%
\bibitem [{\citenamefont {{Zeldovich}}\ and\ \citenamefont
  {{Sunyaev}}(1969)}]{1969Ap&SS...4..301Z}%
  \BibitemOpen
  \bibfield  {author} {\bibinfo {author} {\bibfnamefont {Y.~B.}\ \bibnamefont
  {{Zeldovich}}}\ and\ \bibinfo {author} {\bibfnamefont {R.~A.}\ \bibnamefont
  {{Sunyaev}}},\ }\href {\doibase 10.1007/BF00661821} {\bibfield  {journal}
  {\bibinfo  {journal} {Astrophys.Space.Sci.}\ }\textbf {\bibinfo {volume}
  {4}},\ \bibinfo {pages} {301} (\bibinfo {year} {1969})}\BibitemShut {NoStop}%
\bibitem [{\citenamefont {Sunyaev}\ and\ \citenamefont
  {Zeldovich}(1972)}]{Sunyaev:1972eq}%
  \BibitemOpen
  \bibfield  {author} {\bibinfo {author} {\bibfnamefont {R.}~\bibnamefont
  {Sunyaev}}\ and\ \bibinfo {author} {\bibfnamefont {Y.}~\bibnamefont
  {Zeldovich}},\ }\href@noop {} {\bibfield  {journal} {\bibinfo  {journal}
  {Comments Astrophys. Space Phys.}\ }\textbf {\bibinfo {volume} {4}},\
  \bibinfo {pages} {173} (\bibinfo {year} {1972})}\BibitemShut {NoStop}%
\bibitem [{\citenamefont {Refregier}\ \emph {et~al.}(2000)\citenamefont
  {Refregier}, \citenamefont {Komatsu}, \citenamefont {Spergel},\ and\
  \citenamefont {Pen}}]{Refregier:2000xz}%
  \BibitemOpen
  \bibfield  {author} {\bibinfo {author} {\bibfnamefont {A.}~\bibnamefont
  {Refregier}}, \bibinfo {author} {\bibfnamefont {E.}~\bibnamefont {Komatsu}},
  \bibinfo {author} {\bibfnamefont {D.~N.}\ \bibnamefont {Spergel}}, \ and\
  \bibinfo {author} {\bibfnamefont {U.-L.}\ \bibnamefont {Pen}},\ }\href
  {\doibase 10.1103/PhysRevD.61.123001} {\bibfield  {journal} {\bibinfo
  {journal} {Phys.Rev.}\ }\textbf {\bibinfo {volume} {D61}},\ \bibinfo {pages}
  {123001} (\bibinfo {year} {2000})},\ \Eprint
  {http://arxiv.org/abs/astro-ph/9912180} {arXiv:astro-ph/9912180 [astro-ph]}
  \BibitemShut {NoStop}%
\bibitem [{\citenamefont {Silk}(1968)}]{Silk:1967kq}%
  \BibitemOpen
  \bibfield  {author} {\bibinfo {author} {\bibfnamefont {J.}~\bibnamefont
  {Silk}},\ }\href {\doibase 10.1086/149449} {\bibfield  {journal} {\bibinfo
  {journal} {Astrophys.J.}\ }\textbf {\bibinfo {volume} {151}},\ \bibinfo
  {pages} {459} (\bibinfo {year} {1968})}\BibitemShut {NoStop}%
\bibitem [{\citenamefont {{Kaiser}}(1983)}]{1983MNRAS.202.1169K}%
  \BibitemOpen
  \bibfield  {author} {\bibinfo {author} {\bibfnamefont {N.}~\bibnamefont
  {{Kaiser}}},\ }\href@noop {} {\bibfield  {journal} {\bibinfo  {journal}
  {Mon.Not.R.Astron.Soc.}\ }\textbf {\bibinfo {volume} {202}},\ \bibinfo
  {pages} {1169} (\bibinfo {year} {1983})}\BibitemShut {NoStop}%
\bibitem [{\citenamefont {Weinberg}(2008)}]{weinberg:COS}%
  \BibitemOpen
  \bibfield  {author} {\bibinfo {author} {\bibfnamefont {S.}~\bibnamefont
  {Weinberg}},\ }\href@noop {} {\emph {\bibinfo {title} {Cosmology}}}\
  (\bibinfo  {publisher} {Oxford University Press},\ \bibinfo {address}
  {Oxford, UK},\ \bibinfo {year} {2008})\BibitemShut {NoStop}%
\bibitem [{\citenamefont {Chluba}\ \emph
  {et~al.}(2012{\natexlab{a}})\citenamefont {Chluba}, \citenamefont {Khatri},\
  and\ \citenamefont {Sunyaev}}]{Chluba:2012gq}%
  \BibitemOpen
  \bibfield  {author} {\bibinfo {author} {\bibfnamefont {J.}~\bibnamefont
  {Chluba}}, \bibinfo {author} {\bibfnamefont {R.}~\bibnamefont {Khatri}}, \
  and\ \bibinfo {author} {\bibfnamefont {R.~A.}\ \bibnamefont {Sunyaev}},\
  }\href@noop {} {\  (\bibinfo {year} {2012}{\natexlab{a}})},\ \Eprint
  {http://arxiv.org/abs/1202.0057} {arXiv:1202.0057 [astro-ph.CO]} \BibitemShut
  {NoStop}%
\bibitem [{\citenamefont {Hu}\ \emph {et~al.}(1994)\citenamefont {Hu},
  \citenamefont {Scott},\ and\ \citenamefont {Silk}}]{Hu:1994bz}%
  \BibitemOpen
  \bibfield  {author} {\bibinfo {author} {\bibfnamefont {W.}~\bibnamefont
  {Hu}}, \bibinfo {author} {\bibfnamefont {D.}~\bibnamefont {Scott}}, \ and\
  \bibinfo {author} {\bibfnamefont {J.}~\bibnamefont {Silk}},\ }\href@noop {}
  {\bibfield  {journal} {\bibinfo  {journal} {Astrophys.J.}\ }\textbf {\bibinfo
  {volume} {430}},\ \bibinfo {pages} {L5} (\bibinfo {year} {1994})},\ \Eprint
  {http://arxiv.org/abs/astro-ph/9402045} {arXiv:astro-ph/9402045 [astro-ph]}
  \BibitemShut {NoStop}%
\bibitem [{\citenamefont {Chluba}\ and\ \citenamefont
  {Sunyaev}(2011)}]{Chluba:2011hw}%
  \BibitemOpen
  \bibfield  {author} {\bibinfo {author} {\bibfnamefont {J.}~\bibnamefont
  {Chluba}}\ and\ \bibinfo {author} {\bibfnamefont {R.}~\bibnamefont
  {Sunyaev}},\ }\href@noop {} {\  (\bibinfo {year} {2011})},\ \Eprint
  {http://arxiv.org/abs/1109.6552} {arXiv:1109.6552 [astro-ph.CO]} \BibitemShut
  {NoStop}%
\bibitem [{\citenamefont {Khatri}\ \emph {et~al.}(2012)\citenamefont {Khatri},
  \citenamefont {Sunyaev},\ and\ \citenamefont {Chluba}}]{Khatri:2011aj}%
  \BibitemOpen
  \bibfield  {author} {\bibinfo {author} {\bibfnamefont {R.}~\bibnamefont
  {Khatri}}, \bibinfo {author} {\bibfnamefont {R.~A.}\ \bibnamefont {Sunyaev}},
  \ and\ \bibinfo {author} {\bibfnamefont {J.}~\bibnamefont {Chluba}},\
  }\href@noop {} {\bibfield  {journal} {\bibinfo  {journal}
  {Astron.Astrophys.}\ }\textbf {\bibinfo {volume} {540}},\ \bibinfo {pages}
  {A124} (\bibinfo {year} {2012})},\ \Eprint {http://arxiv.org/abs/1110.0475}
  {arXiv:1110.0475 [astro-ph.CO]} \BibitemShut {NoStop}%
\bibitem [{\citenamefont {Chluba}\ \emph
  {et~al.}(2012{\natexlab{b}})\citenamefont {Chluba}, \citenamefont
  {Erickcek},\ and\ \citenamefont {Ben-Dayan}}]{Chluba:2012we}%
  \BibitemOpen
  \bibfield  {author} {\bibinfo {author} {\bibfnamefont {J.}~\bibnamefont
  {Chluba}}, \bibinfo {author} {\bibfnamefont {A.~L.}\ \bibnamefont
  {Erickcek}}, \ and\ \bibinfo {author} {\bibfnamefont {I.}~\bibnamefont
  {Ben-Dayan}},\ }\href@noop {} {\  (\bibinfo {year} {2012}{\natexlab{b}})},\
  \Eprint {http://arxiv.org/abs/1203.2681} {arXiv:1203.2681 [astro-ph.CO]}
  \BibitemShut {NoStop}%
\bibitem [{Note1()}]{Note1}%
  \BibitemOpen
  \bibinfo {note} {A code for calculating the radiation transfer function is
  available at {\protect \sf http://www.mpa-garching.mpg.de/\textasciitilde
  {}komatsu/CRL/}. This code is based on CMBFAST \cite
  {Seljak:1996is}}\BibitemShut {NoStop}%
\bibitem [{\citenamefont {Kogut}\ \emph {et~al.}(2011)\citenamefont {Kogut},
  \citenamefont {Fixsen}, \citenamefont {Chuss}, \citenamefont {Dotson},
  \citenamefont {Dwek} \emph {et~al.}}]{Kogut:2011xw}%
  \BibitemOpen
  \bibfield  {author} {\bibinfo {author} {\bibfnamefont {A.}~\bibnamefont
  {Kogut}}, \bibinfo {author} {\bibfnamefont {D.}~\bibnamefont {Fixsen}},
  \bibinfo {author} {\bibfnamefont {D.}~\bibnamefont {Chuss}}, \bibinfo
  {author} {\bibfnamefont {J.}~\bibnamefont {Dotson}}, \bibinfo {author}
  {\bibfnamefont {E.}~\bibnamefont {Dwek}},  \emph {et~al.},\ }\href {\doibase
  10.1088/1475-7516/2011/07/025} {\bibfield  {journal} {\bibinfo  {journal}
  {JCAP}\ }\textbf {\bibinfo {volume} {1107}},\ \bibinfo {pages} {025}
  (\bibinfo {year} {2011})},\ \Eprint {http://arxiv.org/abs/1105.2044}
  {arXiv:1105.2044 [astro-ph.CO]} \BibitemShut {NoStop}%
\bibitem [{Note2()}]{Note2}%
  \BibitemOpen
  \bibinfo {note} {{\protect \sf http://cmbpol.kek.jp/litebird/}}\BibitemShut
  {NoStop}%
\bibitem [{\citenamefont {{The Planck
  Collaboration}}(2006)}]{2006astro.ph..4069T}%
  \BibitemOpen
  \bibfield  {author} {\bibinfo {author} {\bibnamefont {{The Planck
  Collaboration}}},\ }\href@noop {} {\bibfield  {journal} {\bibinfo  {journal}
  {ArXiv Astrophysics e-prints}\ } (\bibinfo {year} {2006})},\ \Eprint
  {http://arxiv.org/abs/arXiv:astro-ph/0604069} {arXiv:astro-ph/0604069}
  \BibitemShut {NoStop}%
\bibitem [{\citenamefont {Ade}\ \emph {et~al.}(2011)\citenamefont {Ade} \emph
  {et~al.}}]{PlanckHFICoreTeam:2011az}%
  \BibitemOpen
  \bibfield  {author} {\bibinfo {author} {\bibfnamefont {P.}~\bibnamefont
  {Ade}} \emph {et~al.} (\bibinfo {collaboration} {Planck HFI Core Team}),\
  }\href@noop {} {\  (\bibinfo {year} {2011})},\ \Eprint
  {http://arxiv.org/abs/1101.2039} {arXiv:1101.2039 [astro-ph.IM]} \BibitemShut
  {NoStop}%
\bibitem [{\citenamefont {Hinshaw}\ \emph {et~al.}(2009)\citenamefont {Hinshaw}
  \emph {et~al.}}]{Hinshaw:2008kr}%
  \BibitemOpen
  \bibfield  {author} {\bibinfo {author} {\bibfnamefont {G.}~\bibnamefont
  {Hinshaw}} \emph {et~al.} (\bibinfo {collaboration} {WMAP Collaboration}),\
  }\href {\doibase 10.1088/0067-0049/180/2/225} {\bibfield  {journal} {\bibinfo
   {journal} {Astrophys.J.Suppl.}\ }\textbf {\bibinfo {volume} {180}},\
  \bibinfo {pages} {225} (\bibinfo {year} {2009})},\ \Eprint
  {http://arxiv.org/abs/0803.0732} {arXiv:0803.0732 [astro-ph]} \BibitemShut
  {NoStop}%
\bibitem [{\citenamefont {Jarosik}\ \emph {et~al.}(2011)\citenamefont {Jarosik}
  \emph {et~al.}}]{Jarosik:2010iu}%
  \BibitemOpen
  \bibfield  {author} {\bibinfo {author} {\bibfnamefont {N.}~\bibnamefont
  {Jarosik}} \emph {et~al.},\ }\href {\doibase 10.1088/0067-0049/192/2/14}
  {\bibfield  {journal} {\bibinfo  {journal} {Astrophys.J.Suppl.}\ }\textbf
  {\bibinfo {volume} {192}},\ \bibinfo {pages} {14} (\bibinfo {year} {2011})},\
  \Eprint {http://arxiv.org/abs/1001.4744} {arXiv:1001.4744 [astro-ph.CO]}
  \BibitemShut {NoStop}%
\bibitem [{\citenamefont {Agullo}\ and\ \citenamefont
  {Shandera}(2012)}]{agullo-shandera}%
  \BibitemOpen
  \bibfield  {author} {\bibinfo {author} {\bibfnamefont {I.}~\bibnamefont
  {Agullo}}\ and\ \bibinfo {author} {\bibfnamefont {S.}~\bibnamefont
  {Shandera}},\ }\href@noop {} {\  (\bibinfo {year} {2012})},\ \Eprint
  {http://arxiv.org/abs/1204.4409} {arXiv:1204.4409 [astro-ph.CO]} \BibitemShut
  {NoStop}%
\bibitem [{\citenamefont {Seljak}\ and\ \citenamefont
  {Zaldarriaga}(1996)}]{Seljak:1996is}%
  \BibitemOpen
  \bibfield  {author} {\bibinfo {author} {\bibfnamefont {U.}~\bibnamefont
  {Seljak}}\ and\ \bibinfo {author} {\bibfnamefont {M.}~\bibnamefont
  {Zaldarriaga}},\ }\href {\doibase 10.1086/177793} {\bibfield  {journal}
  {\bibinfo  {journal} {Astrophys.J.}\ }\textbf {\bibinfo {volume} {469}},\
  \bibinfo {pages} {437} (\bibinfo {year} {1996})},\ \Eprint
  {http://arxiv.org/abs/astro-ph/9603033} {arXiv:astro-ph/9603033 [astro-ph]}
  \BibitemShut {NoStop}%
\end{thebibliography}
\end{document}